\documentclass[trackchanges,twocolumn]{aastex701} 
\usepackage{amsmath}
\usepackage{float}
\usepackage{xcolor}

\usepackage{booktabs}

\newcommand{\Halpha}{H{$\alpha$}}
\newcommand{\Halphaline}{H{$\alpha$}$\lambda$6563}

\newcommand{\MgII}{Mg{\sc ~ii}}

\newcommand{\NII}{N{\sc ~ii}}
\newcommand{\NIIline}{[N{\sc ~ii}$]\lambda$6583}

\newcommand{\OVI}{O{\sc ~vi}}
\newcommand{\OII}{O{\sc ~ii}}

\newcommand{\OIII}{O{\sc ~iii}}
\newcommand{\OIIIline}{[O{\sc ~iii}$]\lambda$5007}
\newcommand{\OIIdoubline}{[O{\sc ~ii}$]\lambda\lambda$3727, 3729}

\newcommand{\sbunits}{$\rm{erg\,s^{-1}\, cm^{-2}\, arcsec^{-2}}$}

\newcommand{\HI}{H{\sc ~i}}
\newcommand{\HII}{H{\sc ~ii}}

\newcommand{\HeI}{He{\sc ~i}}
\newcommand{\HeII}{He{\sc ~ii}}
\newcommand{\HeIII}{He{\sc ~iii}}

\newcommand{\pynbody}{\texttt{PYNBODY}}
\newcommand{\changa}{{\sc ChaNGa}}
\newcommand{\cloudy}{\texttt{CLOUDY}}
\newcommand{\Mstar}{$M_\star$}

\newcommand{\Rvir}{$R_{\rm vir}$}
\newcommand{\rHI}{$R_{\rm HI}$}

\newcommand{\HalpSB}{H$\alpha$ SB}
\newcommand{\NIISB}{[N{\sc ~ii}] SB}
\newcommand{\OIIISB}{[O{\sc ~iii}] SB}

\newcommand{\HalpSBres}{$\Delta \langle$H$\alpha$$\rangle$}
\newcommand{\NIISBres}{$\Delta \langle$[N{\sc ~ii}]$\rangle$}
\newcommand{\OIIISBres}{$\Delta \langle$[O{\sc ~iii}]$\rangle$}

\begin{document}


\title{The CGM in a New Light: Emitted \Halpha, [\NII], and [\OIII] around Simulated Dwarf Galaxies}

\author[orcid=0000-0003-0200-6986]{Daniel R. Piacitelli}
\affiliation{Rutgers University, Department of Physics and Astronomy, Piscataway, NJ 08854, USA}\email[show]{piacitelli.danielr@gmail.com}

\author{Alyson M. Brooks}
\affiliation{Rutgers University, Department of Physics and Astronomy, Piscataway, NJ 08854, USA}\affiliation{Center for Computational Astrophysics, Flatiron Institute, 162 Fifth Ave, New York, NY 10010, USA}\email{abrooks@physics.rutgers.edu}

\author[0000-0001-7589-6188]{N. Nicole Sanchez}\affiliation{The Observatories of the Carnegie Institution for Science, 813 Santa Barbara Street, Pasadena, CA 91101, USA}\affiliation{Cahill Center for Astronomy and Astrophysics, California Institute of Technology, MC249-17, Pasadena, CA 91125, USA}\email{nnicolesanchez@gmail.com}

\author[0000-0001-6779-3429]{Charlotte Christensen}\affiliation{Physics Department, Grinnell College, 1116 Eighth Avenue, Grinnell, IA 50112, USA }\email{christenc@grinnell.edu}

\author[orcid=0000-0002-6558-9894]{Chloe Neufeld} 
\affiliation{Department of Astronomy, Yale University, 219 Prospect Street, New Haven, CT 06511, USA}
\email{chloe.neufeld@yale.edu}

\author[orcid=0000-0002-7075-9931]{Imad Pasha}
\affiliation{Department of Astronomy, Yale University, 219 Prospect Street, New Haven, CT 06511, USA}\email{}
\affiliation{Dragonfly Focused Research Organization, 150 Washington Avenue, Suite 201, Santa Fe, NM 87501, USA}

\author[orcid=0000-0002-8282-9888]{Pieter van Dokkum}
\affiliation{Department of Astronomy, Yale University, 219 Prospect Street, New Haven, CT 06511, USA}\email{}
\affiliation{Dragonfly Focused Research Organization, 150 Washington Avenue, Suite 201, Santa Fe, NM 87501, USA}

\author[0000-0001-7831-4892]{Akaxia Cruz}
\affiliation{Center for Computational Astrophysics, Flatiron Institute, 162 Fifth Ave, New York, NY 10010, USA}\affiliation{Department of Physics, Princeton University, Princeton, NJ 08544, USA}\affiliation{Department of Astrophysical Sciences, Princeton University, Princeton, NJ 08544, USA}\email{acruz@flatironinstitute.org}

\author[0000-0002-9642-7193]{Ben Keller}
\affiliation{Department of Physics and Materials Science, University of Memphis, 3720 Alumni Avenue, Memphis, TN 38152, USA}\email{bkeller1@memphis.edu}

\author[0000-0001-5510-2803]{Thomas R. Quinn}
\affiliation{Department of Astronomy, University of Washington, Seattle, WA 98195, USA}\email{trq@astro.washington.edu}

\author{Sijing Shen}\affiliation{Institute of Theoretical Astrophysics, University of Oslo, PO Box 1029, Blindern 0315, Oslo, Norway}\email{sijing.shen@astro.uio.no}

\author{James Wadsley}
\affiliation{Department of Physics \& Astronomy, McMaster University, ABB-241, 1280 Main Street West, Hamilton, Ontario, L8S 4M1, Canada}\email{wadsley@mcmaster.ca}

\begin{abstract}
Mapping the circumgalactic medium (CGM) in emission is a critical next step in understanding its role in the baryon cycle and galaxy evolution, and dwarf galaxies ($M_{\star}\lesssim 10^9 M_{\odot}$) are essential laboratories to study the effects of feedback on the CGM. Using the Marvelous Massive Dwarfs simulation suite, we model the CGM surface brightness (SB) around dwarf galaxies in \Halphaline, \NIIline, and \OIIIline, lines accessible to current integral-field spectrographs (e.g., MUSE and KCWI) and to narrow-band imaging with the MOTHRA telephoto array. We investigate trends in SB magnitude, extent, and geometry as well as correlations with gas kinematics, recent star formation, and galaxy size. Emission in the three lines typically traces distinct CGM structures and phases. SB and the size of the emitting region can differ significantly at a given galaxy mass, but enhancements or suppression in either brightness or extent correlate well across lines. Typically, emission is preferentially eccentric and oriented along the galaxy's semi-minor axis. 
While outflows are key drivers in dispersing metals into the CGM and shaping SB, they remain difficult to identify in SB maps and contribute $\sim40\%$ of the total SB across lines and galaxy masses, while inflows contribute $\sim50\%$. Finally, we identify evidence for a positive correlation between SB and recent star formation and weaker but still significant evidence for a negative correlation between SB and \HI\, disk size. These results link CGM emission to both a galaxy’s recent star formation and its cumulative evolution, yielding predictions that ongoing surveys can test and further establishing emission mapping as a new window into the baryon cycle.
\end{abstract}


\section{Introduction} 
The circumgalactic medium (CGM) is the diffuse, multiphase atmosphere within which galaxies reside. The CGM holds the majority of a galaxy's baryonic budget \citep{Tumlinson17, Piacitelli25} and mediates galaxy evolution as the interface between the interstellar medium (ISM) and intergalactic medium. Galaxies across the mass spectrum depend on inflows from the CGM to fuel star formation  \citep{Tumlinson17}, while feedback processes can generate outflows that distribute energy, mass, and metals back into the CGM \citep[e.g.,][]{Heckman90}. Therefore, mapping the CGM is critical to understanding how galaxies grow and in turn shape their environment via feedback processes.

Dwarf galaxies ($M_{\star} \lesssim 10^9 \rm\, M_{\odot}$) are essential laboratories for studying feedback due to their shallower gravitational potential wells \citep{Somerville_Dave15}. Observationally, the imprint of this feedback sensitivity has been noted in the declining metallicities along the mass-metallicity relation \citep{Tremonti04,Ma16} and CGM \OVI\, detections extending to twice the size of the halo \citep{Johnson_17, Tchernyshyov_22, Mishra_24,Dutta_25_col}, which suggest that feedback-driven outflows escape dwarfs more readily than higher-mass galaxies, evacuating metals from the disk. As the repository of these outflows, the dwarf CGM is critical to understanding dwarf galaxy evolution and the role of feedback across the mass spectrum.

Up to now, the overwhelming majority of CGM observational programs have used absorption spectroscopy to detect the CGM around a wide range of galaxy masses  \citep[e.g.,][]{Bordoloi_14,Werk16, Burchett19, Chen20, Zheng2024, Mishra_24} and derive integrated column densities in a variety of ions. These studies have yielded insights regarding the extent of the metal-enriched CGM and the complex and multiphase nature of the CGM. However, absorption-based CGM measurements face key limitations: (1) the availability of quasars directly constrains the number of galaxies for which the CGM is observed, (2) in most cases, only a single quasar sightline is available for a given halo, limiting insights into the global CGM structure, and (3) column densities integrate over gas at a range of densities and temperatures, obscuring the information on local conditions.

Emission observations will map a larger fraction of the CGM than absorption along sightlines and can reveal substructure, such as accreting filaments or outflow signatures. A growing number of observational surveys have begun mapping the CGM in emission. Integral field unit (IFU) spectroscopy with the Keck Cosmic Web Imager (KCWI) has detected CGM emission in \Halpha, $\rm H\beta$, \OIIdoubline, and \OIIIline\ out to $\sim 30\rm\, kpc$ from the host galaxy. These observations have constrained the ionization mechanisms operating at the disk–CGM interface \citep[${\text \Mstar} = 10^{10}\,M_{\odot}$;][]{Nielsen24} and the total CGM gas mass \citep[${\text \Mstar} = 10^{9.2}\, M_{\odot}$;][]{Lin25}. Beyond KCWI, \citet{2026arXiv260718385Z} use MaNGA IFU data to link CGM emission profiles to galaxy star formation rates. Additionally, by stacking Multi Unit Spectroscopic Explorer (MUSE) data from around ${\text \Mstar}=10^{6} - 10^{12}M_{\odot}$ galaxies, \citet{Dutta24_musestack} find extended [\OII], [\OIII], and $\rm H\beta$ emission and constrain both the ionization parameter and metallicity gradients. 

In addition to current instruments, MOTHRA (Modular Optical Hyperspectral Robotic Array), a new instrument building on the legacy of the Dragonfly Array \citep{Abraham_2014} and Dragonfly Spectral Line Mapper \citep{Lokhorst24, Pasha24, Chen25}, began commissioning in fall 2026. Furthermore, the Aspera smallsat \citep{Chung21b} is set to launch in January 2027 and will target \OVI\, emission, and the planned Habitable Worlds Observatory is also poised to target CGM emission \citep[HWO;][]{HWOCase}. Ultimately, emission mapping is the cornerstone of the next phase of CGM observations, and modern simulations are needed to inform observations at this new frontier.

Prior theoretical work on CGM emission has primarily focused on UV emission from $z\geq1$ Milky Way-mass halos \citep[e.g.,][]{Corlies20,Nelson21, Lochhaas26, Saeedzadeh26}. Critically, no theoretical work addresses the optical emission from low-$z$ dwarf CGM that MOTHRA is currently targeting. This work investigates how the surface brightness (SB) in the MOTHRA-targeted lines trends with galaxy masses and how it may correlate with galaxy properties. 

This paper is organized as follows: Section \ref{sec:sim} describes the simulation suite used, emissivity modeling, and galaxy selection criteria. Section \ref{sec:sbmaps} presents the simulated SB maps and analyzes the extent and geometry of emission. Section \ref{disc:gasprop} determines the thermodynamic conditions and metallicities traced by emission. Section \ref{disc:kinematic} quantifies how inflows and outflows contribute to the total SB.  Section \ref{disc:corr} presents correlations between the SB in each line and galaxy properties. Section \ref{sec:disc} discusses the limitations of our study, and we summarize our conclusions in Section \ref{sec:summary}.

\section{Simulations}\label{sec:sim}
This work uses the Marvelous Massive Dwarfs (MMD) simulation suite \citep{Cruz_25} ran with the smoothed particle hydrodynamic (SPH) code \changa\, \citep{CHANGA}. MMD comprises $41$ individual $25^3\text{ Mpc}^3$ zoom-in volumes that are centered on dwarf galaxies selected from the {\sc Romulus25} cosmological volume simulation \citep{Tremmel17} and utilized in \citet{Keith2025, Ruan2025, Piacitelli25, Cruz_25, Geda26, Piacitelli26}. Initial conditions were generated using \textsf{GenetIC} \citep{GenetIC2021}, and the volumes are run within a Planck cosmology \citep{Planck16cosmo}. MMD has a force resolution of 87 pc, gas particle mass resolution of 3300 $M_{\odot}$, initial star particle mass resolution of 994 $M_{\odot}$, and dark matter particle mass resolution of 17,900 $M_{\odot}$.

Dark matter halos are identified using the Amiga Halo Finder \citep{AHF_09}. The open-source analysis package, \pynbody\, \citep{pynbody}, is used for particle selection and analysis of the \changa\, simulations.

Metal line cooling is modeled across temperatures $10-10^{9}\rm\,K$, interpolating cooling and heating rates from \cloudy\, tables based on the particle density, temperature, metallicity, and redshift, and assuming photoionization equilibrium under the \citet{HM2012} redshift-dependent UV background \citep[see][for more details]{Shen2010}. The total radiative cooling in the MMD simulations also incorporates cooling/heating from Compton and free-free processes and cooling from primordial species (\HI, H$_2$, \HII, \HeI, \HeII, \HeIII) as well as prescriptions for H$_2$ self-shielding and dust shielding of \HI\ \citep{Christensen2012}.

Star formation in the simulations is restricted to high-density regions where H$_2$ is present, following the prescription in \citet{Christensen2012}.  Star particles are born with a \citet{Kroupa2001_IMF} initial mass function.  Mass is returned to the interstellar medium via stellar winds, supernovae (SNe) Ia, and SNe II.  Each SN II injects $10^{51}$ erg of thermal energy into the surrounding gas particles, and the energy propagates following the superbubble model of \citet{Keller14}.  The simulations also include the formation and growth of black holes (BHs). BH formation is based on local gas conditions, and therefore the model makes no assumptions about occupation fraction. More specifically, BH particles (with a seed mass of $10^{5}$ $\rm M_{\odot}$) form when the gas particles are in dense ($n_{\rm H} > 1.5 \times 10^4\ \rm cm^{-3}$), low metallicity  ($Z < 10 ^{-5}$), cool gas ($T < 5 \times 10^3$ K) with a large Jeans mass ($M_J > 10^5$ M$_{\odot}$) and low H$_2$ fraction ($X_{\rm H_2} < 2 \times 10^{-3}$) \citep{Bellovary2010}.  These conditions are generally only met at high redshift.

BH growth follows a boosted Bondi model \citep{Booth09,Tremmel17}. To compensate for limited resolution, accretion is boosted above a threshold density $n_{\rm th}$, and accretion energy couples to surrounding gas with efficiency $e_{\rm fb}$. The fiducial model adopts $n_{\rm th}=100\,\mathrm{cm^{-3}}$ and $e_{\rm fb}=0.005$.


\subsection{Emissivity Calculations}

In post-processing, the line volume emissivity, $\epsilon_V$, for a given particle is tabulated based on the particle's hydrogen number density, temperature, metallicity, and redshift ($n_{\rm H}, T, Z, z$) using \cloudy\, \citep{Cloudy_Ferland17}. These \cloudy\, tables model ionization equilibrium of both photoionization and collisional ionization processes in the presence of the \citet{HM2012} UV background. The 4D grid of $\epsilon_V$ produced by \cloudy\, has the following parameter space: $-7.0 \leq \log(n_{\rm H}/\rm cm^{-3}) \leq 5.0$ ($\Delta\,\log(n_{\rm H})=0.10$), $1.0 \leq \log(T /K) \leq 8.0$ ($\Delta\,\log(T)=0.10$), $-4.0 \leq \log(Z/Z_{\odot}) \leq 2.0$ ($\Delta\,\log(Z/Z_{\odot})=0.50$), and $0.0 \leq z \leq 1.0$ ($\Delta\,z=0.50$). However, in this study, we only select the $z=0$ snapshots.  

We then linearly interpolate this table for the entire simulation volume to derive $\epsilon_V$ for each gas particle. Line luminosities are then calculated by multiplying $\epsilon_V$ by the particle volume $V$, which we derive based on the quotient of the particle mass and density ($V=m/\rho$), and the gas-phase species fraction relative to solar. MMD tracks the oxygen mass fraction ($f_O$) for each particle explicitly. Nitrogen mass fraction ($f_N$) is found during the analysis phase by scaling the particle oxygen abundance by the relative abundance of oxygen to nitrogen ($A_N/A_O$) using  $f_N = f_O \times (A_N/A_O)$. We adopt $A_O=8.4 \times 10^{-3}$ and $A_N=9.2 \times 10^{-4}$ values following \citet{Asplund2009}.

MMD models SN-heated particles as two-phase particles. These two phases maintain pressure equilibrium and have distinct masses, densities, and temperatures. Rather than placing the particles in either phase or averaging the temperatures, we model the emissivity of each phase separately and then take the volume-weighted average $\epsilon_V$ when making the SB maps. This method is self-consistent with the simulation modeling schemes and conserves each particle's total line luminosity.

To produce SB maps, we use \pynbody\, to take all gas particles within a given box size around a galaxy and project the gas volume emissivity $\epsilon_V$ across each pixel's line of sight. With these projected maps, we multiply by the pixel area and divide by the distance factor ($4\pi D^2$) and the solid angle per pixel ($\Omega$) to derive an SB map. Throughout this work, we generate $500 \times 500$ pixel SB maps with a side length of $3{\text \Rvir}$ (i.e., spanning $\pm1.5\,{\text \Rvir}$ from the galaxy center). Because the maps scale with each halo's virial radius, the physical pixel size varies across the sample. We verified that our results are unchanged when we instead adopt a fixed physical side length of $200\rm\, kpc$ for all galaxies, indicating that either choice is sufficient to resolve trends in SB. 

In this work, we do not model photon attenuation or absorption along the line of sight. While the CGM in metal lines is typically optically thin, \Halpha\, emission, particularly in the disk region, will likely be affected by attenuation. Thus, the predictions for \Halpha\, emission are likely upper limits, and future work producing mock SB maps should consider attenuation effects in more depth. Nonetheless, here we focus on the extended SB around these galaxies and investigate trends in SB magnitude, extent, and geometry as well as potential correlations with the baryon cycle, star formation, and galaxy size. 

\subsection{Galaxy Selection}

\begin{figure}
    \centering
    \includegraphics[width=0.9\linewidth]{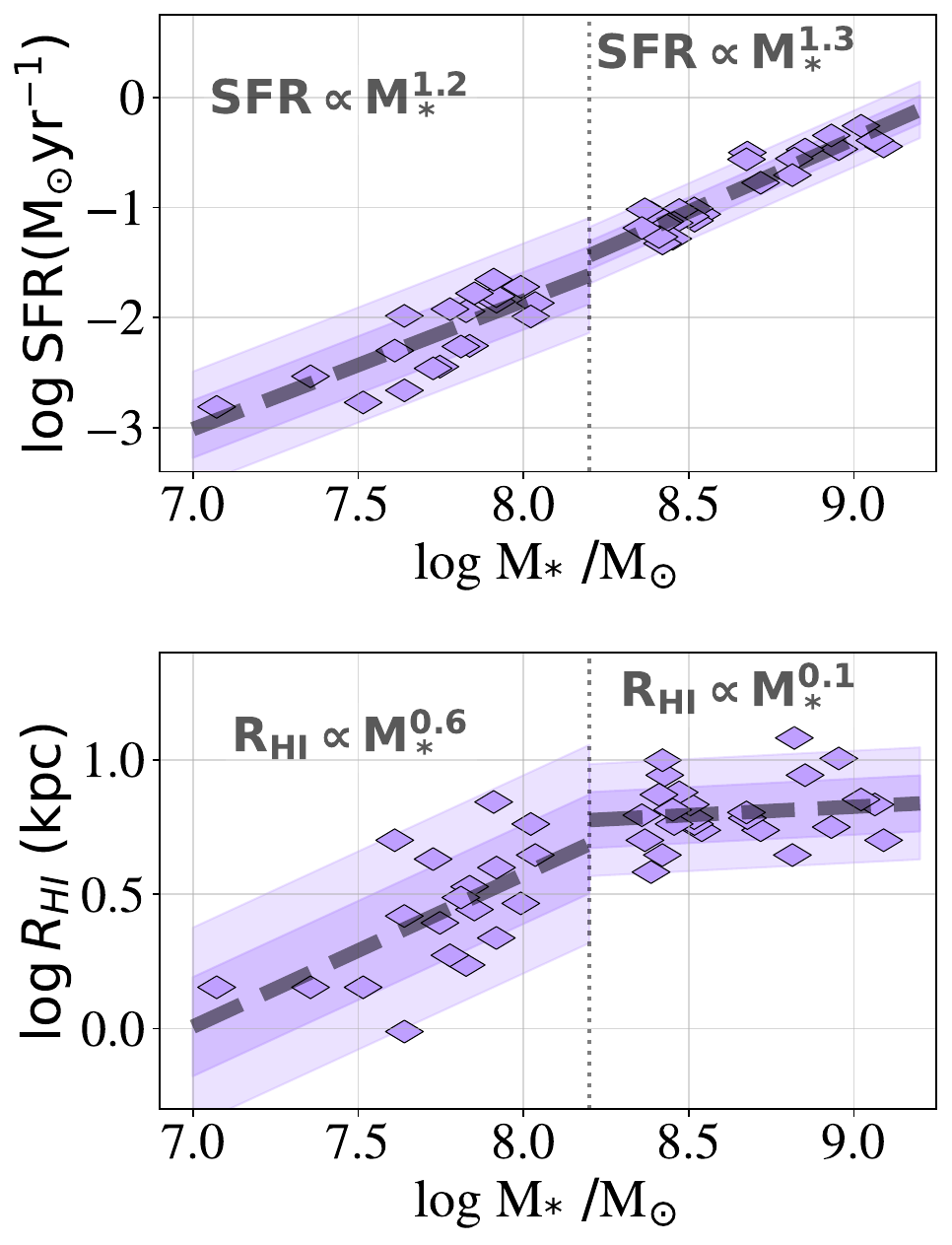}
    \caption{Star formation rates (SFR) measured over a $100\, \rm{Myr}$ timescale (top panel) and \HI\, disk size above a surface density of $1 \rm\ M_{\odot} \ pc^{-2}$ (\rHI; bottom panel) plotted in log-log space. SFR and \rHI\, values are fit with a broken power law shown as grey dashed lines and broken at $\log M_{\star}/M_{\odot}=8.2$ (vertical grey dotted line). We chose the break mass based on the gap in galaxies included in MMD at these masses and the apparent steepening in the slope of \rHI\ below this mass. Shaded bands represent the $1\sigma$ and $2\sigma$ scatter about the fit lines.}
    \label{fig:intro}
\end{figure}

\begin{figure*}
    \centering
    \includegraphics[width=1\linewidth]{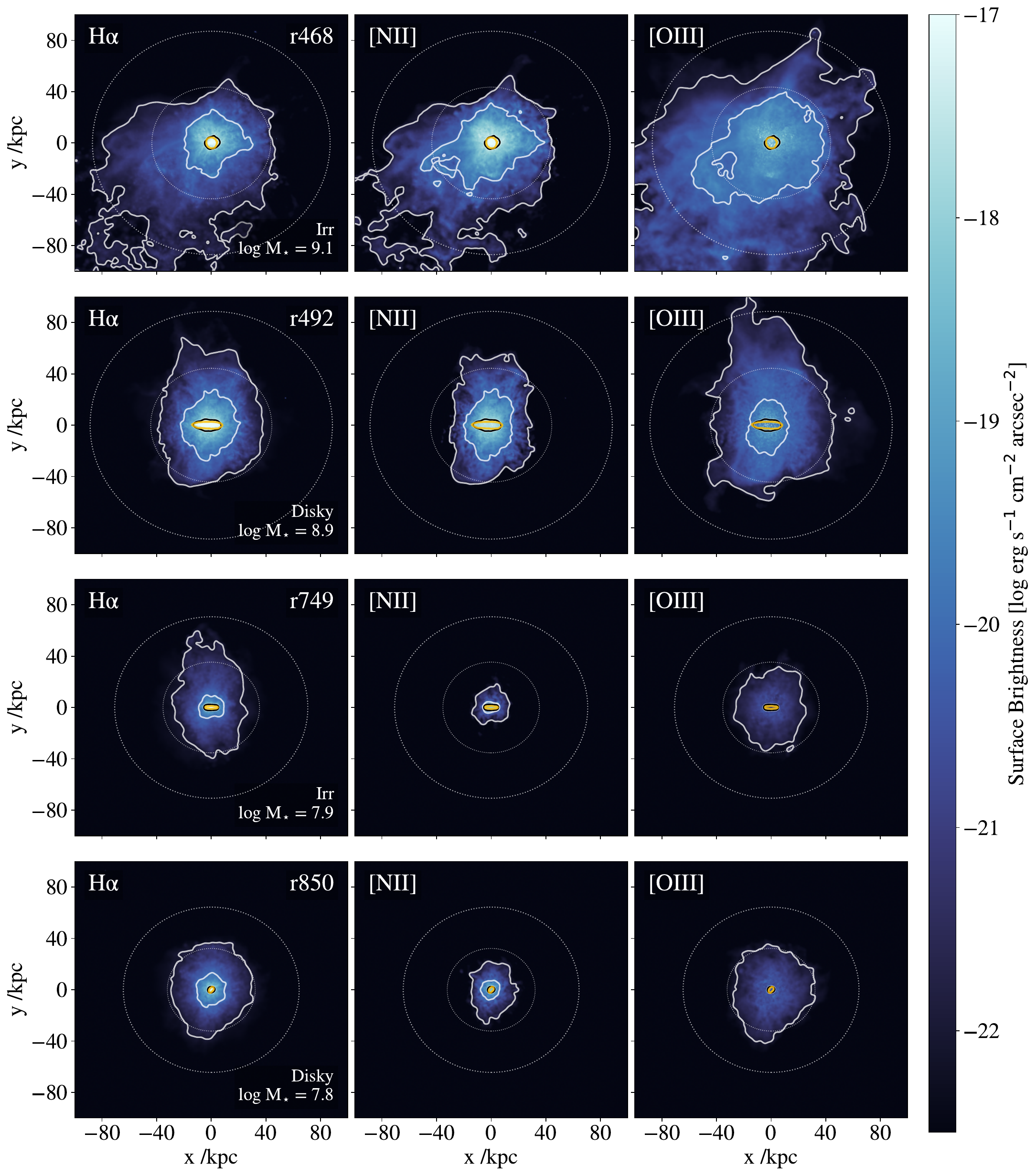}
    \caption{Surface brightness (SB) maps in \Halpha\, (left), [\NII] (center), and [\OIII] (right) for four representative galaxies at $z=0$, oriented edge-on based on the angular momentum vector. The top two rows show two $\log M_{\star}/M_{\odot}\sim 9$ galaxies and the bottom two rows show two representative $\log M_{\star}/M_{\odot}\sim 8$ galaxies; both mass bins include a disky and non-disky galaxy to showcase how galaxy morphology may shape CGM SB.  For reference, we include contours describing the size of the disk/halo: the $B$-band stellar SB brighter than $\mu_B=27.5 \rm\, mag~ arcsec^{-2}$ (yellow), HI surface density above $\Sigma_{\rm HI}=1\rm\, M_{\odot} ~pc^{-2}$ (black), and $0.5\ R_{\rm vir}$ and $1\ R_{\rm vir}$ (dotted white circles) are overplotted on each map. The solid white contours are SB contours at $10^{-20}$~\sbunits and $10^{-22}$~\sbunits, the approximate expected threshold for MOTHRA observations is $2\times 10^{-21}$~\sbunits. We find that observable SB in all three lines extends well into the CGM, with each line tracing differing structures.}
    \label{fig:sbmaps}
\end{figure*}

We select galaxies from MMD within a stellar mass range of $10^7\leq M_{\star} \leq 10^9 ~M_{\odot}$ at the $z=0$ snapshot to match the planned MOTHRA observations of local dwarf galaxies. All simulated stellar masses are taken to be $0.6$ times the total star-particle mass, as this is a better match to photometric stellar mass determinations \citep{Munshi13}. An isolation criterion is also imposed to remove the effects of satellite galaxies from our analysis by only selecting galaxies that are isolated within $90\rm\, kpc$ from another $ M_{\star} \geq 10^7 ~M_{\odot}$ galaxy.


Using these criteria, we select 44 galaxies from MMD. Galaxies are classified as either ``Irregular'' or ``Disky'' following the same methods as \citet{Geda26}. We kinematically classify galaxies as disks if the median specific angular momentum of their stars in the $\hat{z}$ direction exceeds $0.5$, after aligning the galaxy's total angular momentum vector in $\hat{z}$ based on the gas in the inner $5$ kpc.

Figure \ref{fig:intro} presents the recent star formation rates (SFR) and \HI\, disk sizes (\rHI) of the selected galaxies. SFR values are derived over the last $100 ~\rm{Myr}$ \citep[identical method to Figure 1 of][]{Geda26}. Specific SFRs can be found in Figure 1 of \citet{Geda26} and Figure 2 of \citet{Piacitelli25}. \rHI\, values are the average radius at which the \HI\, surface density falls below $1 ~\rm M_{\odot} ~pc^{-2}$. SFR values decline with declining stellar mass, while disk size weakly declines for $\rm M_{\star}> 10^{8.2} ~M_{\odot}$ galaxies, and disk size declines much more rapidly for $\rm M_{\star}< 10^{8.2} ~M_{\odot}$ galaxies.

In order to describe the phenomenological relationship between SFR, disk size, and stellar mass, we fit broken power laws to each data set, using the same stellar mass for the break point in each fit. More specifically, broken power-law fits are plotted on each panel for $\rm M_{\star}>10^{8.2} ~M_{\odot}$ (hereafter, high-mass galaxies) and $\rm M_{\star}<10^{8.2} ~M_{\odot}$ (hereafter, low-mass galaxies). In addition to the best-fit lines, the shaded bands represent the $1\sigma$ and $2\sigma$ ranges of galaxy scatter about the fit.

The mass at which we break the power-law ($\rm M_{\star}=10^{8.2} ~M_{\odot}$) was based on the general trends of galaxies above and below this mass in \rHI\, in particular. Because MMD contains no galaxies in the $10^{8.0}<\rm M_{\star}<10^{8.4} ~M_{\odot}$ range, we adopt simply the midpoint of this gap ($\rm M_{\star}=10^{8.2} ~M_{\odot}$) as the break mass for the fits. A similar break in slope has been shown in the observed stellar-to-atomic gas mass relation \citep{Huang12,Bradford15}. However, observed breaks tend to fall within $10^{8.6}\leq{\text \Mstar} /\rm\, M_{\odot} \leq 10^{9}$, notably higher masses than our choice here. While MMD shows a break above and below this mass, the lack of galaxies prevents a strong determination of whether this mass is truly significant. Future work with more complete coverage in this mass range should investigate the true turnover mass in more detail. Nonetheless, the difference in slopes for the high- and low-mass galaxies motivates analyzing the sample as two mass bins (defined above), and we repeat this fitting procedure throughout this paper.


\section{Synthetic Surface Brightness Maps}\label{sec:sbmaps}

Figure \ref{fig:sbmaps} presents the SB maps in \Halpha\,  (left column), \NIIline\, (center column), and \OIIIline\, (right column) for two $\text{\Mstar}\sim10^9 M_{\odot}$ galaxies and two $\text{\Mstar}\sim10^8 M_{\odot}$ galaxies. For both mass bins, we present an example irregular and disky galaxy. We include contours denoting the $B$-band stellar SB $\mu_B=27.5 \rm\, mag~ arcsec^{-2}$ (yellow), the \HI\, surface density $\Sigma_{\rm HI}=1 \rm\, M_{\odot}~ pc^{-2}$ (black), and $0.5R_{\rm vir}$ and $1.0R_{\rm vir}$ (white dotted). We note that the colorbar for Figure \ref{fig:sbmaps} extends fainter than MOTHRA’s predicted detection limit. 
For reference, we include SB contours (solid white lines) which denote SB thresholds of $10^{-20}\,$\sbunits\, and $10^{-22}\,$\sbunits. These SB contours bracket the expected MOTHRA SB sensitivity of $2\times10^{-21}\,$\sbunits. Distinct differences are apparent in the SB maps across ions and galaxies, even for galaxies of similar masses.

\begin{figure*}
    \centering
    \includegraphics[width=0.98\linewidth]{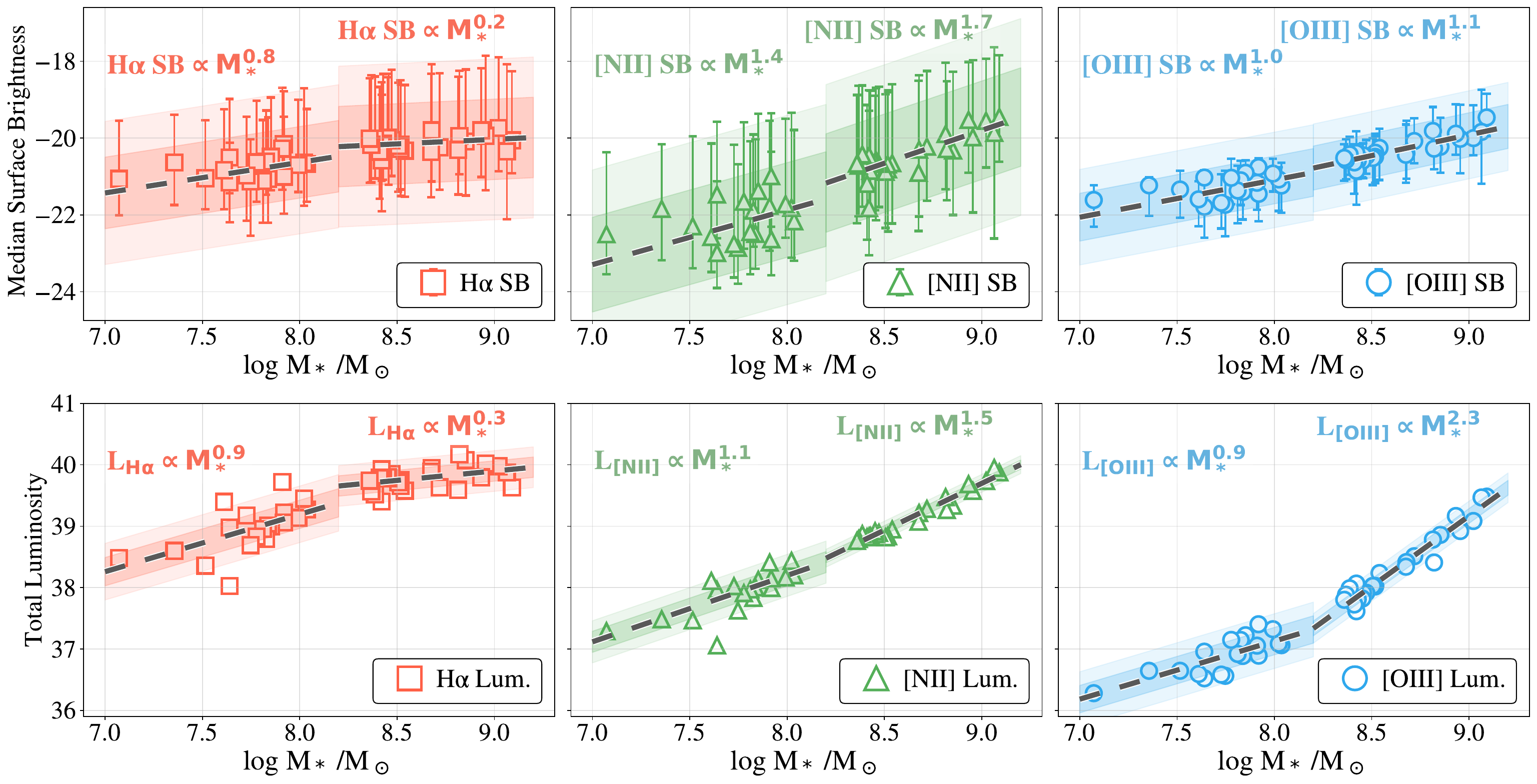}
    \caption{\textit{Top row:} Median $\log\, \rm SB$ each galaxy within $0.5 R_{\rm vir}$ in \Halpha\, (left), \NIIline\, (center), and \OIIIline\, (right). Error bars denote $16-84\rm th$ percentile ranges for the SB values in each map. \textit{Bottom row:} Total luminosity (in $\log \rm erg\ s^{-1}$) in each line for particles within a spherical region with radius $r=0.5 R_{\rm vir}$. We plot power-law fits of $\log \rm{SB}\,$ for the high-mass and low-mass samples and provide the slopes in each panel. Shaded regions correspond to the $1\sigma$ and $2\sigma$ scatter of the computed $\log \rm{SB}$ about the fit line. Generally, the brightness of metal lines falls off more steeply with declining stellar mass than \Halpha. 
    }
    \label{fig:Avg_SB}
\end{figure*}

\textbf{\Halpha\ SB:} Across galaxy masses, \Halpha\, emission extends well beyond the disk of the galaxy and into the halo, typically reaching the $0.5R_{\rm vir}$ boundary and, in some cases (e.g., r468), the $1.0R_{\rm vir}$ boundary. For high-mass galaxies, the \Halpha\, SB appears to trace both dense and clumpy/filamentary gas as well as misty and volume-filling material further away from the galaxy. With declining stellar mass, the contribution from this clumpy/filamentary gas is less apparent in the SB maps, likely due to both declining numerical resolution (i.e., fewer particles) as well as these galaxies having smaller dense, cold gas reservoirs due to their masses. 

\textbf{\NIIline\, SB:} The \NIIline\, SB is strongest near the disk, and the [\NII]-emitting region covers a similar extent to \Halpha\, for high-mass galaxies, but a smaller extent for low-mass galaxies. This difference likely reflects the stronger gas-phase metallicity dependence, and therefore \Mstar, of \NIISB\ than \HalpSB\, and will be explored in the following sections. Like \Halpha, [\NII] emission primarily comes from more filamentary gas, and similar filamentary structures appear in both SB maps, suggesting that the emission in these lines can often co-spatial. 

\textbf{\OIIIline\, SB:} The [\OIII]-emitting region extends well beyond the $\mu_B$ and $\Sigma_{\rm HI}$ boundaries for both high- and low-mass galaxies. Although substructure can be identified in the \OIIISB\, maps, [\OIII] emission appears to preferentially trace a more volume-filling medium across galaxy masses. Notably, the [\OIII]-emitting regions are of substantially different extents between r468 and r492, likely attributable to recent activity in the galaxy.

%
%
%
%
\subsection{Surface Brightness Scaling Relations}\label{subsec:sb_avg}
%

\begin{figure}
    \centering
    \includegraphics[width=0.98\linewidth]{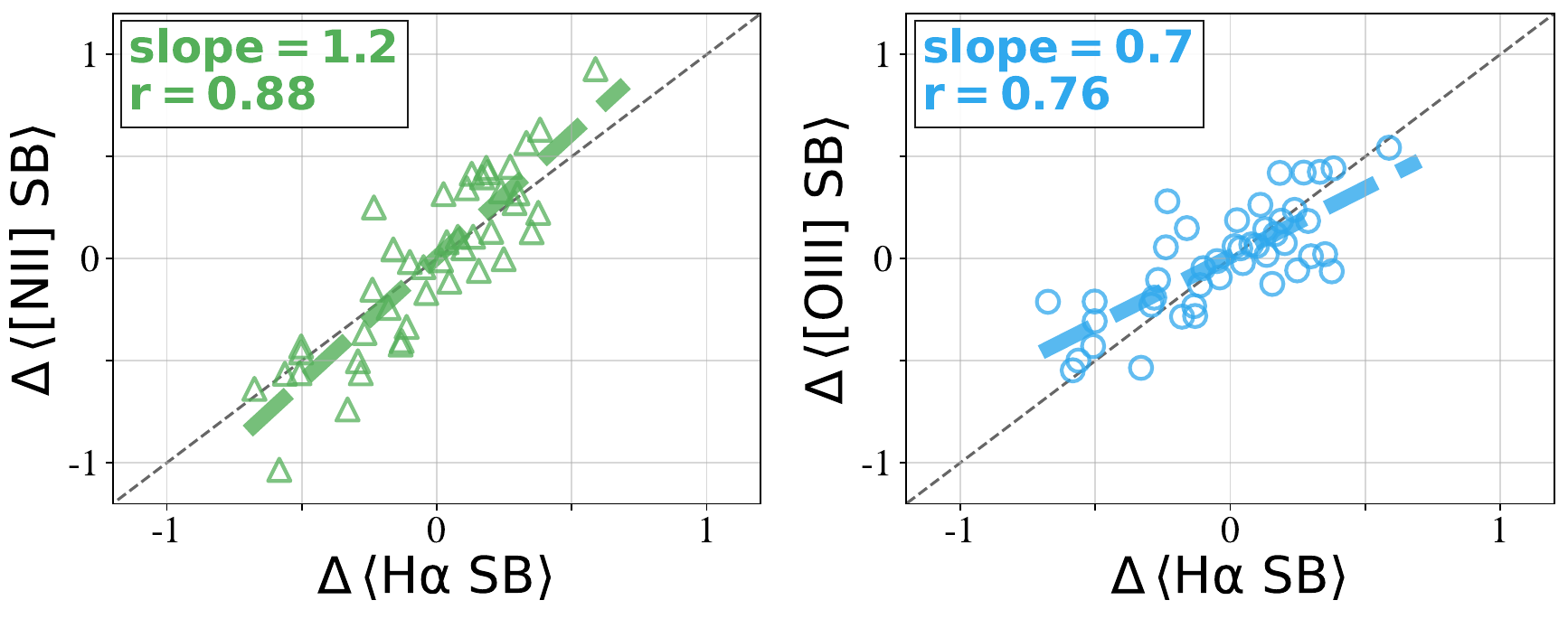}
    \caption{Residuals in median SB values from the fit lines in Figure \ref{fig:Avg_SB}. The horizontal axis for both panels shows the residuals in \HalpSB\, ($\Delta \langle$H$\alpha$ SB$\rangle$; hereafter \HalpSBres), and the left panel's vertical axis shows the residuals in \NIISB\, ($\Delta \langle$[N{\sc ~ii}] SB$\rangle$; hereafter \NIISBres) and the right panel's vertical axis shows the residuals in \OIIISB\, ($\Delta \langle$[O{\sc ~iii}] SB$\rangle$; hereafter \OIIISBres). Linear fits are shown as dashed lines with the fit-line slopes and Pearson correlation coefficients ($r$) provided in each panel. For reference, a $1:1$ line is shown as a grey dashed line. Galaxies that have enhanced/suppressed SB in one line generally have enhanced/suppressed SB in the other lines as well. Differences in the slope of the fit lines are likely due to the lines tracing certain physical conditions (see Section \ref{disc:gasprop} for more details).} 
    \label{fig:sb_residuals}
\end{figure}

The top row of Figure \ref{fig:Avg_SB} shows the SB for each galaxy. We compute the median SB in 50 equal-width radial annuli ($\Delta r = 0.01\,R_{\rm vir}$) out to a projected radius of $0.5\,R_{\rm vir}$ (the innermost white dotted line in Figure~\ref{fig:sbmaps}). Each point in Figure \ref{fig:Avg_SB} shows the median of this radial profile, and the bars span its $16–84\rm\ th$ percentile range. This weights each radius equally rather than each pixel, and the percentile ranges describe the steepness of the SB profile. The bottom row of Figure \ref{fig:Avg_SB} represents the total line luminosity for all particles within a spherical region of $0.5R_{\rm vir}$. We chose the $0.5R_{\rm vir}$ boundary to focus our study on the inner halo across galaxy masses and exclude the outer halo regions where SB is typically $<10^{-22}$\sbunits\,  (Figure \ref{fig:sbmaps}). We explore variations in SB from viewing angle by initially orienting the galaxy edge-on ($0^{\circ}$, as shown in Figure \ref{fig:sbmaps}) and then rotating it about the $y$-axis toward a face-on orientation ($90^{\circ}$), sampling a total of $30$ evenly spaced viewing angles. The markers in Figure \ref{fig:Avg_SB} represent the median value across viewing angles; however, the variations across viewing angles are typically smaller than the plotted markers. Since the median SB shows little variation with viewing angle, we choose to instead plot the $16-84\rm th$ percentile ranges of SB in the edge-on orientation as the error bars in the upper row of Figure \ref{fig:Avg_SB}.

High-mass (${\text \Mstar}>10^{8.2} ~M_{\odot}$) galaxies typically have $\log\,{\text \HalpSB} = -20.1\pm 0.2$, where the error range corresponds to the standard deviation of all galaxies in this mass bin. For the metal lines, we find $\log\,\rm{\text \NIISB} = -20.5 \pm 0.5$ and $\log\,\rm{\text \OIIISB} = -20.3 \pm 0.3$. Notably, these values agree within $1-2\sigma$ of one another. Similarly for total luminosity, we find $\log\,\rm{L_{\rm H\alpha}} = 39.8 \pm 0.2$, $\log\,\rm{L_{[NII]}} =  39.1 \pm 0.4$, and $\log\,\rm{L_{[OIII]}} =  38.3 \pm 0.6$, which agree within $2\sigma$. Both SB and line luminosity decrease as stellar mass decreases.  For lower-mass galaxies (${\text \Mstar}<10^{8.2} ~M_{\odot}$), we find $\log\,\rm{H\alpha \ SB} = -20.8 \pm 0.3$, while the metal lines yield $\log\,\rm{\text \NIISB} = -22.1 \pm 0.5$ and $\log\,\rm{\text \OIIISB} = -21.3 \pm 0.3$. In luminosity, $\log\,\rm{L_{\rm H\alpha}} = 39.0 \pm 0.4$, $\log\,\rm{L_{[NII]}} =  37.9 \pm 0.4$, and  $\log\,\rm{L_{[OIII]}} =  36.9 \pm 0.3$. 

The width of the SB $16-84\rm th$ percentile ranges in Figure \ref{fig:Avg_SB} describes the slope at which SB declines with increasing distance from the galaxy. As seen in Figure \ref{fig:sbmaps}, \NIISB\, declines more steeply with radius than \HalpSB, and therefore typically has the widest percentile ranges. \OIIISB\ has the narrowest percentile ranges, so its emission is typically distributed farther from the disk, with the shallowest radial decline, across galaxy masses.

To better illustrate how SB and total luminosity depend on galaxy mass, we fit the low-mass and high-mass ends with a broken power law, following the methods in Section \ref{sec:sim}. In addition to best-fit lines, we provide the $1\sigma$ and $2\sigma$ ranges of the data as shaded bands. These bands describe both the individual galaxy scatter about the fit (i.e., residuals) and the mean spread in a given galaxy's $\log\,\rm{SB}$, combined in quadrature.

In both SB and luminosities for \Halpha, the high-mass fits have shallow slopes, which steepen at lower masses. For high-mass galaxies, both \HalpSB\, and total luminosities have shallow slopes of $\rm{H\alpha \ SB}\propto M_{\star} ^{0.2}$ and $\rm{L_{\rm H\alpha}}\propto M_{\star} ^{0.3}$, respectively. At lower masses, the slope of \Halpha\, properties steepens compared to the high-mass end. Specifically, the slopes increase to $\rm{H\alpha \ SB}\propto M_{\star} ^{0.8}$ and $\rm{L_{\rm H\alpha}}\propto M_{\star} ^{0.9}$. In Section \ref{disc:corr}, we investigate how the SB values presented here correlate with galaxy properties.

The metal lines exhibit much steeper slopes in both quantities. At high masses, \NIISB\, shows scalings of $\rm{\text \NIISB}\propto M_{\star} ^{1.7}$ and $\rm{L_{[NII]}}\propto M_{\star} ^{1.5}$ and \OIIISB\, shows scalings of $\rm{\text \OIIISB}\propto M_{\star} ^{1.1}$ and $\rm{L_{[OIII]}}\propto M_{\star} ^{2.3}$. At lower masses, the slope of metal-line properties changes, becoming shallower. The slope of [\NII] properties drop to $\rm{\text \NIISB}\propto M_{\star} ^{1.4}$ and $\rm{L_{[NII]}}\propto M_{\star} ^{1.1}$. The slope in \OIIISB\ shows little change ($\rm{\text \OIIISB}\propto M_{\star} ^{1.0}$) compared to higher masses, but the luminosity significantly shallows to $\rm{L_{[OIII]}}\propto M_{\star} ^{0.9}$.

Using the fit lines in Figure \ref{fig:Avg_SB}, we determine each galaxy's residuals about the broken power-law fit. We present how these residuals in each line correlate with one another in Figure \ref{fig:sb_residuals} where the residuals in \HalpSB\, ($\Delta \langle$H$\alpha$ SB$\rangle$; hereafter \HalpSBres) are plotted on the horizontal axis and the vertical axes show either the residuals in \NIISB\, ($\Delta \langle$[N{\sc ~ii}] SB$\rangle$; hereafter \NIISBres) or the residuals in \OIIISB\, ($\Delta \langle$[O{\sc ~iii}] SB$\rangle$; hereafter \OIIISBres). Plotting in this space demonstrates whether the SB enhancement (i.e., positive residual) or suppression (i.e., negative residual) in a given line co-occurs with the SB enhancement or suppression in the other lines. 

Generally, galaxies that have enhanced or suppressed SB in one line also have enhanced or suppressed SB in the other lines. Using linear regression, we derive the slope for each best-fit line and its Pearson correlation coefficient ($r$). In this space, a slope of $1$ indicates that the SB residuals are equal to the \HalpSBres\, values. \NIISBres\, strongly trends with \HalpSBres\, with a slope of $1.2$ and has a high correlation coefficient of $r=0.88$. \OIIISBres\, have a slope of $0.7$ which is shallower and further from the \HalpSBres\, $1:1$ line than \NIISBres\, and correspondingly has a lower correlation coefficient of $r=0.76$, although this value still indicates a strong trend. This difference in how well the two metal lines trend with \HalpSBres\, is driven by \NII\, having a similar ionization potential ($14.5\rm\, eV$) to hydrogen, while \OIII\, has a significantly higher ionization potential ($35.1\rm\, eV$). Different ionization potentials will trace different physical conditions, and we investigate this further in Section \ref{disc:gasprop}. 

\subsubsection{Luminosity Distribution and Efficiency}

The luminosities in Figure \ref{fig:Avg_SB} are summed over particle data within a spherical region, not a projected distance. Given the differences in ionization potential, these ions will exist in distinct physical conditions and likely differing radial extents. We briefly determine the extent at which $90\%$ of the luminosity in each line is enclosed, $r_{90}$, as a function of \rHI. These values are found by creating a cumulative luminosity profile for all particle data within $0.5R_{\rm vir}$. We find that the majority of luminosity in the lower ionization state lines is preferentially tracing the ISM, with  $90\%$ of the \Halpha\, and [\NII] luminosity being enclosed within $1.7\pm 0.7\ {\text \rHI}$ and $2.3\pm 1.0\ {\text \rHI}$. Conversely, [\OIII] luminosity is more extended, with $r_{90}=6.7\pm 3.6\ {\text \rHI}$. 

\begin{figure}
    \centering
    \includegraphics[width=0.9\linewidth]{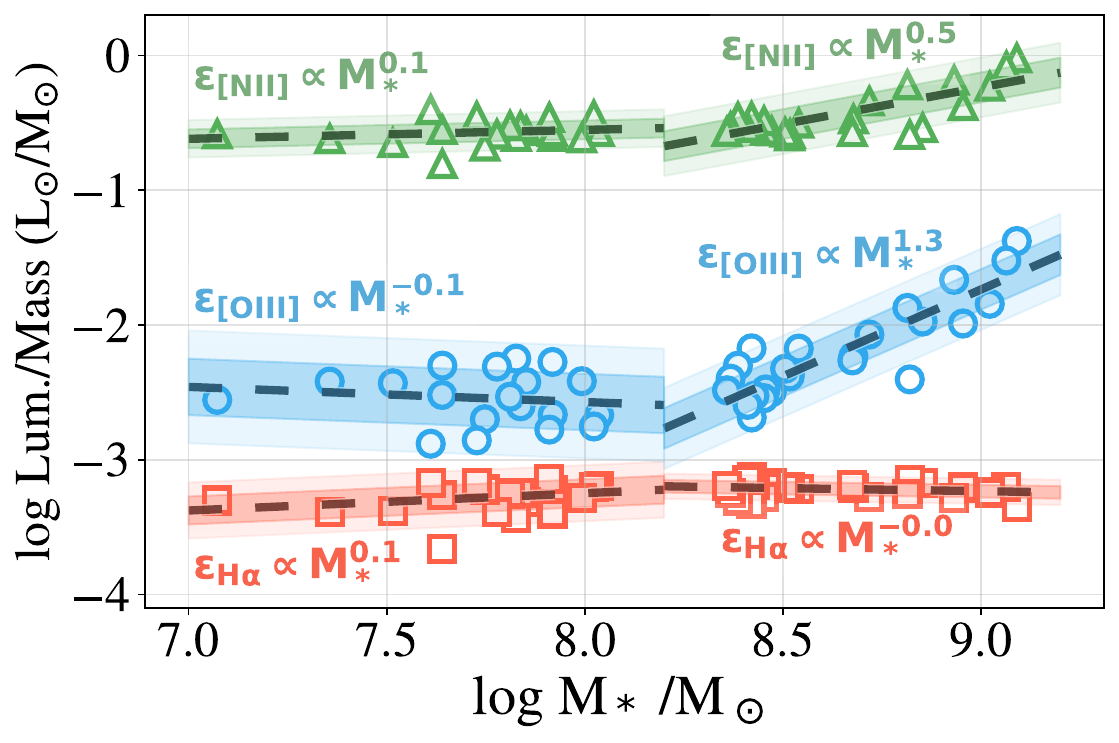}
    \caption{Luminosity efficiency, found by dividing the total luminosity (Figure \ref{fig:Avg_SB}) by the total element mass within $0.5R_{\rm vir}$. Marker styles follow the same presentation as Figure \ref{fig:Avg_SB}. Across stellar masses, [\NII] luminosity is the most efficient per unit mass, whereas \Halpha\ is the least. \Halpha\ shows no strong mass dependence; however, for high-mass galaxies the metal lines become increasingly efficient with increasing mass, likely reflective of the underlying metallicity of the galaxy/CGM (see Figure \ref{fig:rhi_Z}). } 
    \label{fig:Leff}
\end{figure}

Dividing the total luminosity, $L$, by the total element mass, $M_X$, provides a measure of the luminosity efficiency, $\epsilon$, for each line (e.g., $\epsilon_{\text \Halpha} = \sum L_{\text \Halpha} / \sum M_{\rm H}$). Figure \ref{fig:Leff} presents these efficiencies, with luminosities and masses summed within $0.5\,R_{\rm vir}$ and expressed in solar units ($L_\odot/M_\odot$). At all stellar masses, [\NII] is the most efficient line, exceeding the others by $>1 \rm\ dex$. $\epsilon_{\rm H\alpha}$ is typically the least efficient and varies little with stellar mass, with a median value of $\log\epsilon_{\rm H\alpha} = -3.2 \pm 0.1$. At lower masses, the metal-line efficiencies also depend only weakly on mass (slopes of $\propto10^{\pm0.1}$), with $\log\epsilon_{[\rm NII]} = -0.57 \pm 0.09$ and $\log\epsilon_{[\rm OIII]} = -2.53 \pm 0.19$. At higher masses, the metal line luminosity efficiencies steepen with $M_\star$, as seen in the luminosities in Figure \ref{fig:Avg_SB}, reaching $\log\epsilon_{[\rm NII]} = -0.51 \pm 0.16$ and $\log\epsilon_{[\rm OIII]} = -2.30 \pm 0.33$.

\begin{figure*}[!t]
    \centering
    \includegraphics[width=0.98\linewidth]{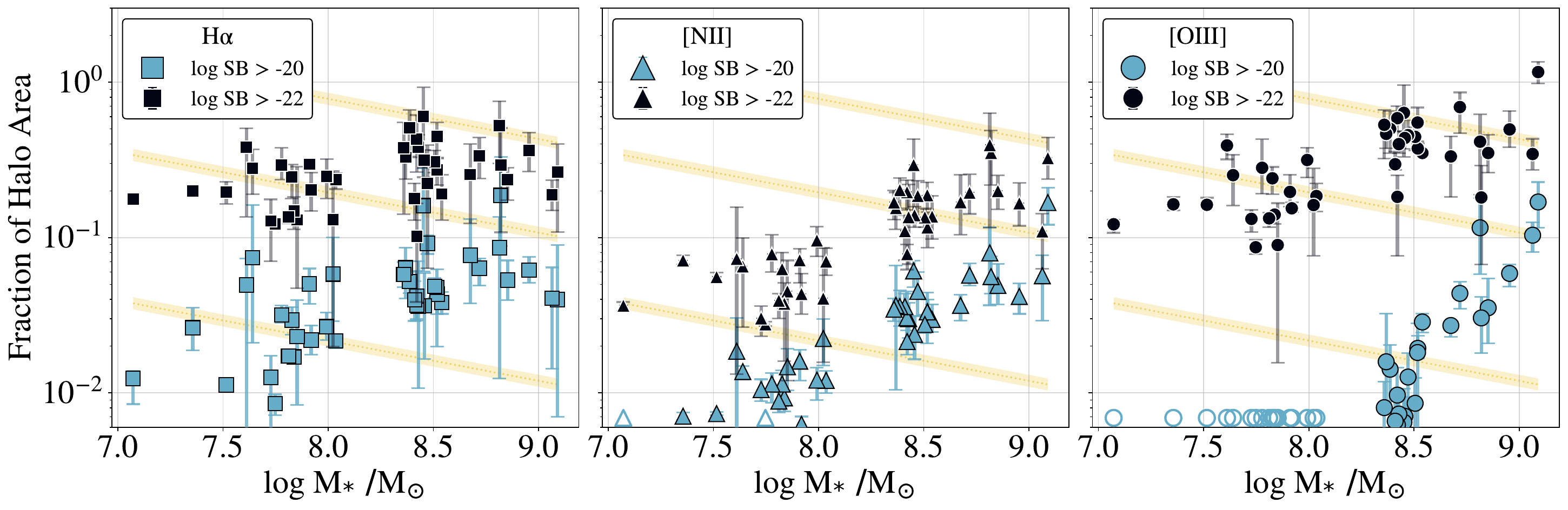}
    \caption{Halo covering fractions above a fixed SB threshold of $10^{-20}$~\sbunits\, (light blue markers) and $10^{-22}$~\sbunits\, (black markers) for \Halpha\, (left), \NIIline\, (center), and \OIIIline\, (right). Covering fractions for each galaxy are shown with respect to stellar mass. We derive these values by calculating the total area enclosed by the SB contour at the chosen threshold, then dividing by the total halo area. To account for viewing-angle variations, we orient the galaxy edge-on and then rotate it from edge-on to face-on through $30$ evenly spaced viewing angles, deriving the covering fraction for each angle. Plotted markers denote the mean covering fraction over all viewing angles, while the error bars represent the minimum to maximum. For reference of physical scale, orange lines and bands are included that represent the fractional size of a $r=10,\ 30,\ 60\rm\ kpc$ halo relative to the mean \Rvir\ at that stellar mass. Galaxies with mean bright covering fractions below $0.6\%$ (2 in [\NII], 17 in [\OIII], all with $M_\star < 10^{8.1} M_{\odot}$) are shown as open-faced upper limits at the bottom of each panel. \Halpha-emitting regions are typically comparable or larger in covering fractions compared to [\NII] and [\OIII] for both SB thresholds. }
    \label{fig:R_SB}
\end{figure*}
\begin{figure}[]
    \centering
    \includegraphics[width=0.98\linewidth]{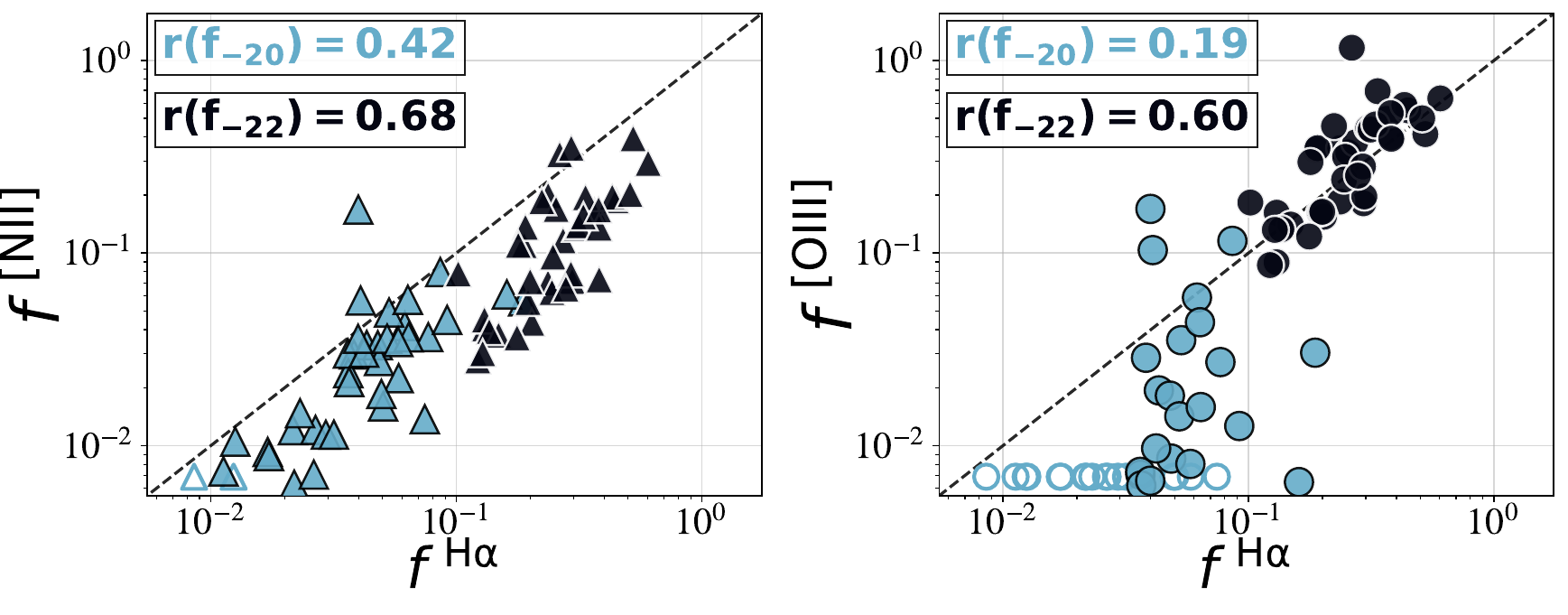}
    \caption{Comparison of SB covering fractions from Figure \ref{fig:R_SB} across lines. Values for \Halpha\, are plotted on the horizontal axis, while [\NII] (left) and [\OIII] (right) are plotted on the vertical axes. Similar to Figure \ref{fig:sb_residuals}, we use Pearson correlation coefficients to quantify how the covering fraction of SB scales across lines. For bright and faint SB in all lines (except bright [\OIII] emission), the extent of emission shows significant correlations indicating that extended emission in one line typically coincides with extended SB in the other lines.}
    \label{fig:R_SB_corr}
\end{figure}

%
%
%
%
\subsection{Surface Brightness Extent}\label{subsec:sb_ext}
%

Figure \ref{fig:R_SB} shows the extent of regions above SB limits of $10^{-20}$ \sbunits\, (hereafter, ``bright'') and $10^{-22}$ \sbunits\, (hereafter, ``faint'') for each simulated galaxy. These values are derived by generating contours on a projected SB map at the given SB limit, then finding the total physical area, $A$, bounded by the contour. Areas are determined with Gauss's area formula (or shoelace formula) from the coordinates of the contour points. These bright and faint SB areas are then divided by the area of the halo ($A_{\rm halo} =  \pi R_{\rm vir}^{2}$) to describe the fraction of the halo covered by this SB limit. We account for variations in SB size from viewing angles using the same method as in Figure \ref{fig:Avg_SB}. SB covering fractions are derived for $30$ evenly spaced viewing angles ranging from edge-on ($0^{\circ}$) to face-on ($90^{\circ}$). Figure \ref{fig:R_SB} plots the mean value across viewing angles, with error bars indicating the minimum and maximum values across all viewing angles. For reference, the orange dotted lines show the fraction of the halo area covered by a circle of fixed physical radius $r = 10$, $30$, and $60\rm\ kpc$ (i.e., $(r/R_{\rm vir})^2$), where \Rvir\ is given by a fit of \Rvir\ against \Mstar\ for our sample. Shaded regions show the $1\sigma$ scatter in \Rvir\ about this fit.

The y-axis of Figure \ref{fig:R_SB} is truncated at $0.6\%$, and galaxies with covering fractions below this value are shown as open-faced upper limits. This cutoff was chosen to remove galaxies with very small emission regions, while still showing the viewing angle variations of higher-mass galaxies with means above this threshold.

Across the three lines, the area covered decreases with declining stellar mass. Splitting the sample by \Mstar, high-mass galaxies have bright SB regions with covering fractions of $f^{\rm H\alpha}_{-20} = 0.06 \pm 0.05$, $f^{\rm [NII]}_{-20} = 0.05 \pm 0.03$, and $f^{\rm [OIII]}_{-20} = 0.04 \pm 0.03$, where error ranges correspond to the combined variance from galaxy-galaxy scatter and the individual galaxy's variance from viewing angles. Across lines, the $f_{-20}$ values lie within $1\sigma$ of each other. Although the mean $f_{-20}$ values are similar for high-mass galaxies, [\NII] and [\OIII] values across the full mass range decline more steeply than \Halpha.

For low-mass galaxies, the bright \Halpha\, regions cover $f^{\rm H\alpha}_{-20} =0.03 \pm 0.03$ of the halo, notably within $1\sigma$ of high-mass $f^{\rm H\alpha}_{-20}$ values. On the other hand, the size of the [\NII] regions decline to $f^{\rm [NII]}_{-20} = 0.01 \pm 0.01$ and [\OIII] falls to less than $f^{\rm [OIII]}_{-20} < 0.006$ (below the axis minimum). For these galaxies, the typical extent of bright SB is more strongly separated, with \Halpha\, and [\NII] being the most extended and consistent within a standard deviation, and [\OIII] being highly confined to small areas.

With deeper SB limits ($>10^{-22}$ \sbunits), the SB covering fractions all increase. For the high-mass sample, faint SB covering fractions are $f^{\rm H\alpha}_{-22} =0.32 \pm 0.19$, $f^{\rm [NII]}_{-22} = 0.19 \pm 0.12$, and $f^{\rm [OIII]}_{-22} = 0.46 \pm 0.24$. Notably, $f^{\rm [OIII]}_{-22}$ tends to be the largest compared to the other two lines, which is driven by the gas phase becoming warmer and more diffuse as distance from these dwarf galaxies increases \citep[e.g.,][]{Piacitelli25}.  Similar to the behavior of bright \Halpha-emitting regions, the low-mass sample shows typical faint \Halpha\, covering fractions within $1\sigma$ ($f^{\rm H\alpha}_{-22} = 0.21 \pm 0.10$) of the high-mass sample. [\NII] and [\OIII] decline more rapidly with decreasing stellar mass to  $f^{\rm [NII]}_{-22} = 0.06 \pm 0.03$ and $f^{\rm [OIII]}_{-22} = 0.19 \pm 0.10$.

Figure \ref{fig:R_SB_corr} plots the bright and faint SB covering fractions for the three lines against one another. Generally, extended emission in one line coincides with extended emission in the other lines (i.e., points move along the $1:1$ line towards the upper right of the plot) for both SB limits. This holds for \Halpha\, and [\NII] (left panel of Figure \ref{fig:R_SB_corr}) that exhibit correlation coefficients of $r=0.42$ and $r=0.68$ for bright and faint SB, respectively. For the vast majority of galaxies, bright and faint \Halpha\ emission is $2.0\pm1.0 \times$ and $2.7 \pm 1.2\times$ larger than [\NII] emission, respectively.

In contrast, bright \Halpha\, and [\OIII] emission show the weakest correlation, with  $r=0.19$. This is likely due to the differences in mass dependence seen in Figure \ref{fig:R_SB}, where $f^{\rm H\alpha}_{-20}$ shows a shallower trend with mass while $f^{\rm [OIII]}_{-20}$ exhibits a steeper shape. Because \OIII\, has a higher ionization potential, this difference in mass dependence likely reflects these lines tracing distinct gas phases and structures. As seen in Figure \ref{fig:sbmaps}, bright \Halpha\, emission tends to trace denser substructure, while bright [\OIII] emission is typically misty and volume-filling (see Section \ref{disc:gasprop} for more details on gas conditions traced by these lines). However, when extending the SB limit to faint SB, Figure \ref{fig:sbmaps} shows that \Halpha\, emission can also trace more misty, diffuse material. This difference is clear in the covering fraction correlations: the correlation between faint \Halpha\, and [\OIII] emission is $r=0.60$, much stronger than its bright counterpart. The extent of faint \Halpha\, and [\OIII] emission tends to be comparable, with a median $f^{\rm H\alpha}_{-22}/f^{\rm [OIII]}_{-22}$ ratio of $0.9\pm0.3 \times$. Typically, the galaxies where $f^{\rm H\alpha}_{-22}<f^{\rm [OIII]}_{-22}$ are higher masses, and correspondingly higher metallicity. For high-mass galaxies, $83\%$ of galaxies ($57\%$ of the full sample) have $f^{\rm H\alpha}_{-22}<f^{\rm [OIII]}_{-22}$. For low-mass galaxies, $f^{\rm H\alpha}_{-22}$ and $f^{\rm [OIII]}_{-22}$ tend to be comparable and only $20\%$ of low-mass galaxies have more extended [\OIII] emission.

\subsection{Surface Brightness Geometry}\label{sec:sb_geom}
\begin{figure*}[!t]
    \centering
    \includegraphics[width=0.98\linewidth]{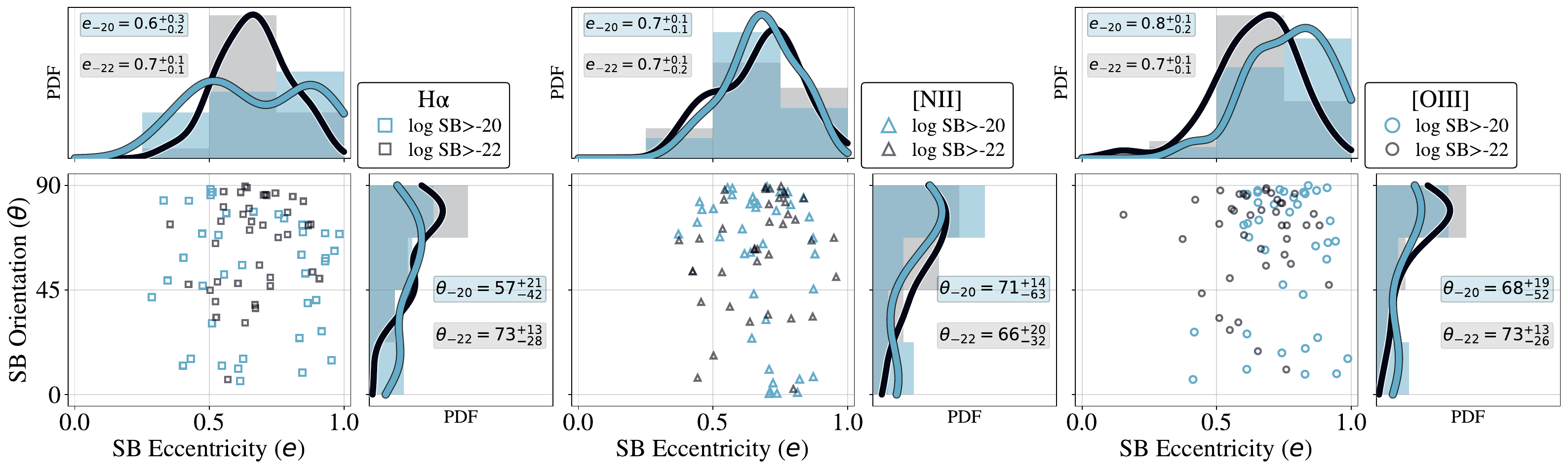}
    \caption{SB geometry diagnostics corner plots. SB contours at limits of $10^{-20}$\sbunits\, (light blue markers) and $10^{-22}$\sbunits\, (black markers) for each line (following the presentation style in Figure \ref{fig:R_SB}) are fit with an ellipse. From the fits, the resulting eccentricity ($e$, horizontal axis) and angle orientation ($\theta$, vertical axis) are derived for each galaxy and plotted. $\theta$ is measured relative to the semi-major axis ($x$ in Figure \ref{fig:sbmaps}). Unlike Figure \ref{fig:R_SB}, only the edge-on oriented SB map is used for this analysis to preserve the physical significance of the semi-minor and semi-major axes. Probability density functions (PDF) for $e$ and $\theta$ are provided, binned in four equally sized bins within $0\leq e\leq1$ and $0\leq \theta \leq 90^{\circ}$; histogram face colors match the SB limit. Solid lines overlaid on the binned data represent the Gaussian kernel-density estimated PDFs. Median values and $16-84\rm th$ percentiles for $e$ and $\theta$ are provided in their respective PDF panels. Generally, bright and faint SB across lines is preferentially aligned with the galaxy's semi-minor axis ($\theta>60^\circ$) and eccentric ($e>0.5$), demonstrating that outflows are likely an important source for CGM SB.}
    \label{fig:SB_shape}
\end{figure*}

In addition to the area spanned by SB in each line, we derive simple geometry diagnostics at both SB limits. Generally, galactic winds are expected to drive metal-enriched, supernova-heated material from the ISM into the CGM, while inflows are expected to be primarily cold, pristine, filamentary gas \citep{Keres05,Brooks09}. In disky galaxies, outflows tend to be preferentially distributed along the semi-minor axis, as feedback-driven material encounters less of the ISM \citep[e.g.,][]{Bland_Tully88}. Signatures of this azimuthal dependence have been noted in observed CGM absorption properties \citep[e.g.,][]{Bordoloi11} as well as in simulations like IllustrisTNG \citep{Peroux20}. However, this conventional wisdom may not hold for dwarf galaxies. In this subsection, we explore the geometry of bright and faint SB.

Using the same SB contours as in Figure \ref{fig:R_SB}, we use \texttt{skimage.measure.regionprops} \citep{scikit} to derive the image eccentricity and orientation. Specifically, we use the SB contours to generate a \texttt{skimage} region within which all pixels are included and weighted equally, and pixels exterior to the contour are excluded. From this region, \texttt{skimage} determines the central moments, $\mu_{pq}$, from the pixel $x$ and $y$ positions following,
\begin{equation}
    \mu_{pq} = \sum_{x} \sum_{y} \biggl(x - \frac{M_{10}}{M_{00}}\biggr)^p ~\biggl(y - \frac{M_{01}}{M_{00}}\biggr)^q ~I(x,y)
\end{equation}
where the raw image moment ratios, $M_{10}/M_{00}$ and $M_{01}/M_{00}$, are defined as $M_{ij}= \sum_{x} \sum_{y} x^i\ y^j \ I(x,y)$, and essentially describe the region centroid \citep{burger2010principles}. $I(x,y)$ is the mask that selects the region within the contour such that $I=0$ for pixels beyond the contour and $I=1$ within. All selected pixels are weighted equally, so each pixel's SB does not affect the image moments; instead, this procedure measures the shape of the region above a SB threshold. From the eigenvectors and eigenvalues of the second-order central moment matrix, the region's eccentricity and orientation are derived. Ultimately, the returned eccentricity and orientation describe an ellipse with the same second spatial moments as the region. When given a perfectly circular region, \texttt{skimage} will return an orientation of $45^{\circ}$; therefore, the orientation angle for low-eccentricity fits is less reliable, although we find that generally all our fits tend to be eccentric. For this analysis, we aim to determine whether SB in certain lines and above certain thresholds is preferentially aligned with the galaxy's semi-major and semi-minor axes, and we perform this image processing only in the edge-on viewing orientation. In this configuration, the $x$ and $y$ axes are aligned with the galaxy's semi-major and semi-minor axes, respectively.

Figure \ref{fig:SB_shape} presents the results of the above analysis. Each line is shown as a corner plot, where the lower-left panel shows the SB eccentricity ($e$) versus the SB orientation ($\theta$), and the upper and right panels show the probability density functions (PDFs) for each parameter. By definition, SB eccentricity ranges from $0\leq e<1$, where $e=0$ describes a circle and higher values correspond to a smaller semi-minor axis relative to the semi-major axis. The orientation angles returned by \texttt{skimage} can fall within the range $\pm 90^{\circ}$. For this analysis, the sign of orientation has less physical significance, so we take the absolute value of the returned angle. Therefore, the range for SB orientation is $0^{\circ}\leq \theta \leq 90^{\circ}$, where $\theta=90^{\circ}$ aligns with the $y$ image direction, or the galaxy semi-minor axis. For each PDF, a Gaussian kernel density estimate is overlaid to mitigate potential biases introduced by binning the data. We see no clear trend in $e$ or $\theta$ with respect to \Mstar. Generally, there is also no clear trend between $e$ and $\theta$ (i.e., highly $e$ fits are also not preferentially high $\theta$). 

Bright and faint SB regions in all three lines tend to be eccentric, $e\geq0.5$, for more than $80\%$ of galaxies. The histograms of eccentricities for both $\log \rm SB>-20$ ($e_{-20}$) and  $\log \rm SB>-22$ ($e_{-22}$) tend to be well-clustered around a central value. Bright \Halpha-emitting regions are less eccentric, with median values of $e_{-20}=0.6$, compared to [\NII] ($e_{-20}=0.7$) and [\OIII] ($e_{-20}=0.8$). However, all median values agree within their $16-84\rm\,th$ percentiles, suggesting that any clear trend is indiscernible in this dataset. Faint SB eccentricities are in stronger agreement across lines, with all three lines having a median eccentricity of $e_{-22}=0.7$.

The fit orientation angles occupy the full $0^{\circ}\leq \theta \leq 90^{\circ}$ range, but there is often a preference for orientations closer to the minor axis. We report the median $\theta\,$ values; however, we emphasize that the $16-84\rm\,th$ percentiles can be large relative to the median.  To better identify general biases in orientation angles, we also define three ranges: ``low-$\theta$'' ($0\leq\theta\leq 30$), ``mid-$\theta$'' ($30^{\circ}\leq\theta\leq 60^{\circ}$), and ``high-$\theta$'' ($60^{\circ}\leq\theta\leq 90^{\circ}$) and report the percent of galaxies in each bin. Physically, the ``low-$\theta$'' range indicates a stronger alignment with the semi-major axis, and the ``high-$\theta$'' range is more aligned with the semi-minor axis.

Bright \Halpha-emitting regions have a fairly uniform distribution in orientation angles, with $\sim30\%$ of galaxies within the low- and mid-$\theta$ ranges and $\sim40\%$ in the high-$\theta$ range, with a median of $\theta_{-20} ({\rm H\alpha}) = 57^{+21}_{-42}$. Still, SB shows a bias toward alignment with the semi-minor axis. Faint \Halpha\ emission is more strongly aligned with the semi-minor axis, with a median value of $\theta_{-22} ({\rm H\alpha}) = 73^{+13}_{-28}$ and $36\%$ and $62\%$ of galaxies in the mid- and high-$\theta$ range, respectively.

Bright and faint [\NII]-emitting regions are also more aligned with the semi-minor axis, with medians of $\theta_{-20} ({\rm [NII]}) = 71^{+14}_{-63}$ and $\theta_{-22} ({\rm [NII]}) = 66^{+20}_{-32}$, and $>65\%$ of galaxies are in the high-$\theta$ range. Similarly, bright and faint [\OIII]-emitting regions have high medians, $\theta_{-20} ({\rm [OIII]}) = 68^{+19}_{-52}$ and $\theta_{-22} ({\rm [OIII]}) = 73^{+13}_{-26}$. Furthermore, $62\%$ and $74\%$ of galaxies have bright and faint orientation angles in the high-$\theta$ range. Thus, for most galaxies, the SB from metal lines tends to be preferentially eccentric and at higher angles, indicative of these lines being primarily sourced by currently or previously feedback-impacted material distributed along the semi-minor axis.

\begin{figure*}
    \centering
    
    \includegraphics[width=0.98\linewidth]{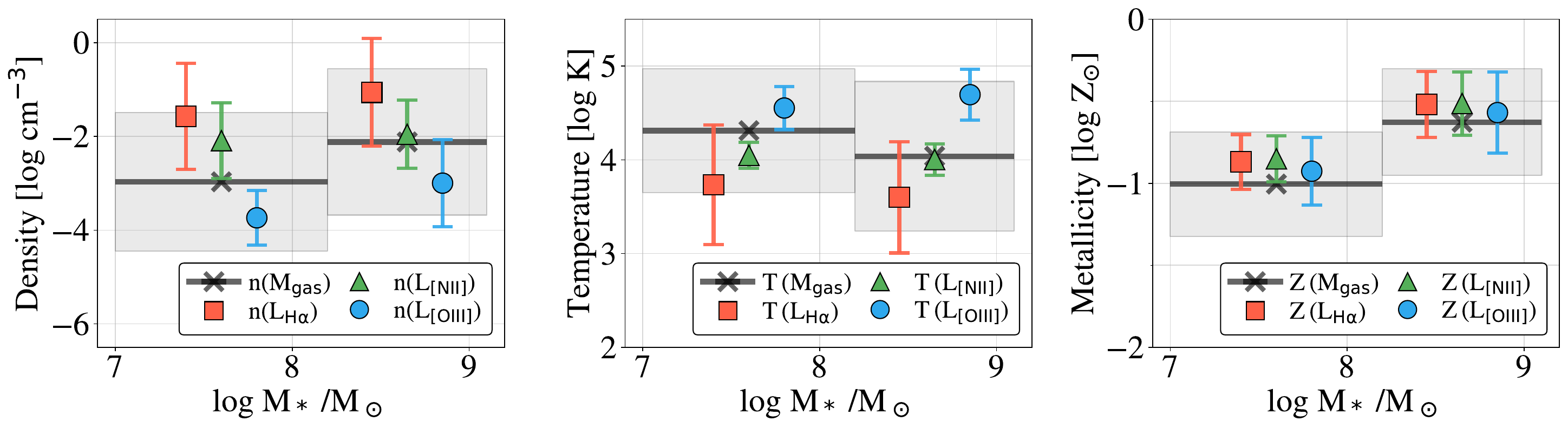}
    \caption{Luminosity-weighted gas density (left), temperature (center), and metallicity (right), binned into low- and high-mass bins. Particle data is selected for all particles within $R<0.5R_{\rm vir}$ (not projected distance). The averages weighted by \Halpha, \NIIline, and \OIIIline\, luminosity are shown in red squares, green triangles, and blue circles, respectively (following the presentation style in Figures \ref{fig:Avg_SB}) and the mass-weighted averages for gas within $R<0.5R_{\rm vir}$ are shown by black cross markers. Error bars incorporate both galaxy-galaxy scatter and individual galaxies' standard deviations. }
    \label{fig:conditons_summ}
\end{figure*}

Despite this strong trend for most galaxies, $20-30\%$ of galaxies sit in the low-$\theta$ range for bright [\NII] and [\OIII]. This is particularly striking because there are only $8\%$ of galaxies in the mid-$\theta$ range, creating a weakly bimodal distribution in orientation angles. Of these low-$\theta$ galaxies, seven are in the low-$\theta$ range for both metal lines. 

There are eight galaxies with bright [\NII] orientation angles of $\theta_{-20}<20^{\circ}$, which are not apparent in the faint SB angles and span the stellar mass range of $10^{7.6}\leq M_{\star}/M_{\odot}\leq 10^{8.6}$. These are disky galaxies which also have less extended bright [\NII] emission at their given masses. As a result, bright \NIISB\, emission traces the disk itself, leading to low $\theta_{-20}$ values and high eccentricities. Although the bright [\NII] emission for these galaxies is well aligned with their semi-major axes, the faint emission is more strongly aligned with their semi-minor axes and falls within $53\leq \theta_{-22}\leq 84^{\circ}$, showcasing that the faint SB may be important to map the history of outflows for some galaxies.

\section{Physical Drivers of Emission}\label{sec:drivers}

The geometry metrics presented in Section \ref{sec:sb_geom} imply that the emitting regions in all three lines are typically elliptical and oriented preferentially along the semi-minor axis of the galaxy. This configuration may arise if outflows or feedback events expel energy and mass into the CGM. This section explores this further by investigating the conditions each line's luminosity traces (Section \ref{disc:gasprop}). We then examine the fraction of SB from inflows, outflows, and static gas in each line (Section \ref{disc:kinematic}) and quantify the correlation between SB and galaxy properties (Section \ref{disc:corr}). 

\subsection{Gas Conditions}\label{disc:gasprop}
The differences in the correlations between \HalpSBres\, and the metal lines seen in Figure \ref{fig:sb_residuals} are likely driven by the three lines tracing different physical conditions. Selecting all particles within $0.5R_{\rm vir}$, we find the weighted average (in log space) for density, temperature, and metallicity. We weight by both the luminosity in the three lines as well as mass.  We present these values in Figure \ref{fig:conditons_summ}, where the sample is broken into low- and high-mass bins. For each stellar mass bin, the mean across galaxies is presented with error bars representing the standard deviations incorporating galaxy-to-galaxy scatter and the standard deviation in each galaxy's particle data.

Across these three conditions, the luminosity-weighted conditions tend to show only a minor mass dependence. In particular, each line's luminosity-weighted conditions for high- and low-mass bins lie within $1\sigma$. Furthermore, the luminosity-weighted conditions all lie within the mass-weighted conditions $1\sigma$ error ranges, indicating that while some lines may trace differing phases, all three trace the bulk structure of the gas.

\textit{Density \&\, Temperature:} In density and temperature space, \Halpha\, has luminosity-weighted conditions of $\bar{n}_{\rm H\alpha} = 10^{-1.3 \pm1.1} \rm cm^{-3}$ and $\bar{T}_{\rm H\alpha} = 10^{3.7 \pm 0.6} \rm K$. With the standard deviations for both $\bar{n}_{\rm H\alpha}$ and $\bar{T}_{\rm H\alpha}$ being the largest compared to the other lines,  \Halpha\, luminosity traces a range of gas phases. While the width of the distributions is wide, the mass-weighted density and temperatures, $\bar{n}_{M}$ and $\bar{T}_{M}$, lie $\sim1\sigma$ from $\bar{n}_{\rm H\alpha}$ and $\bar{T}_{\rm H\alpha}$. Furthermore, for both conditions, the furthest extent of the $\bar{n}_{\rm H\alpha}$ and $\bar{T}_{\rm H\alpha}$ extends beyond the $1\sigma$ range of the mass-weighted conditions. Together, these facts illustrate that \Halpha\, luminosity preferentially traces denser, cooler material.

\begin{figure*}
    \centering
    \includegraphics[width=0.98\linewidth]{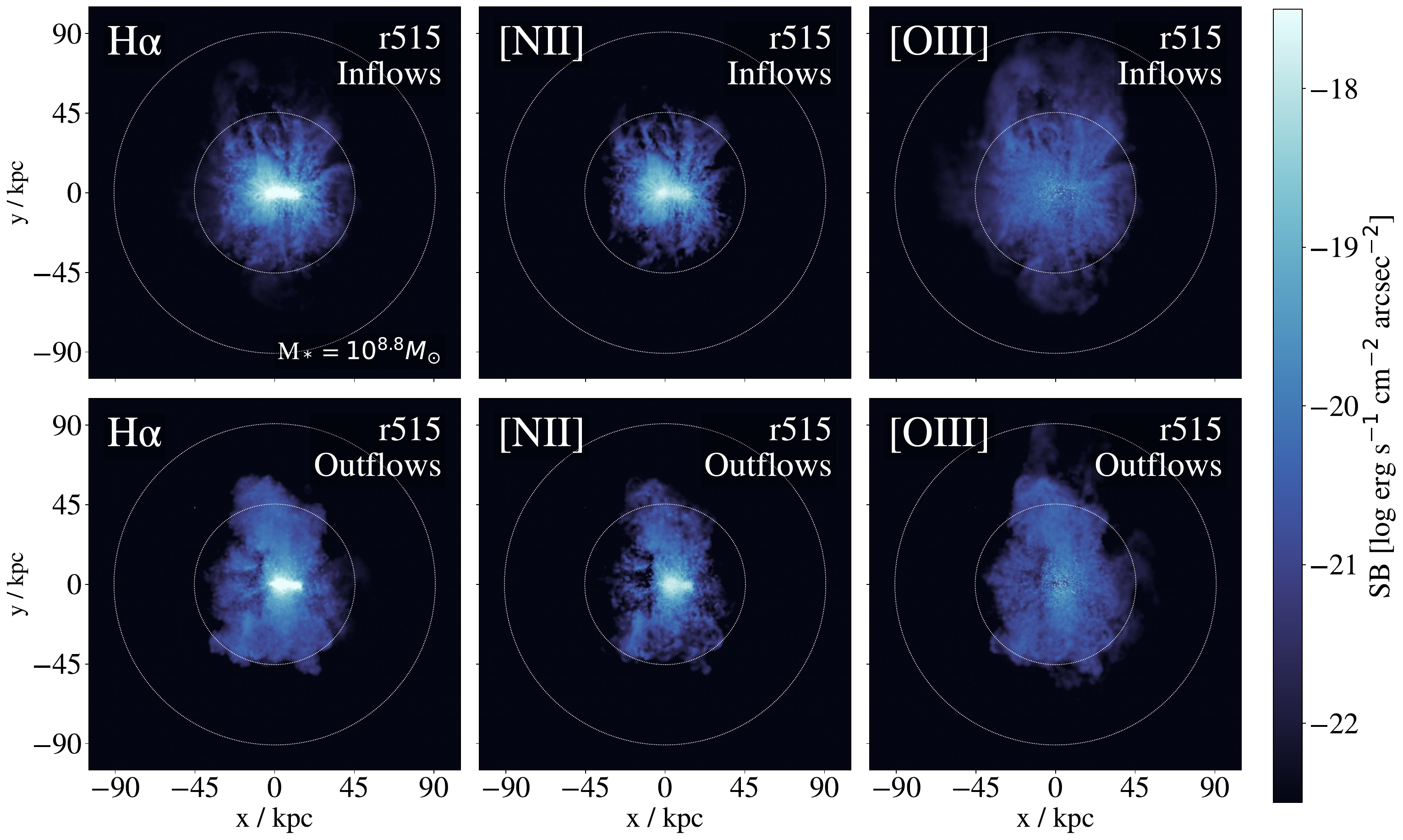}
    \caption{SB maps of a representative $\log M_{\star}/M_{\odot}\sim 9$ galaxy split into inflows (\textit{upper row}) and outflows (\textit{bottom row}). These SB maps are made by first selecting only inflowing ($v_r < -0.1v_{\rm vir}$) and outflowing ($v_r > 0.1v_{\rm vir}$) particles, then using the same SB map pipeline as Figure \ref{fig:sbmaps}. White dotted contours show $0.5 R_{\rm vir}$ and $R_{\rm vir}$.}
    \label{fig:sb_kinesplit}
\end{figure*}

The [\NII] luminosity-weighted density and temperatures tend to be within $1\sigma$ of $\bar{n}_{\rm H\alpha}$ and $\bar{T}_{\rm H\alpha}$. Specifically, $\bar{n}_{\rm [NII]} = 10^{-2.0 \pm 0.8} \rm cm^{-3}$ and  $\bar{T}_{\rm [NII]} = 10^{4.0 \pm 0.1} \rm K$, notably a much narrower temperature range than the \Halpha\, conditions.  In the high-mass bins, $\bar{n}_M$ and $\bar{T}_M$ lie within $1\sigma$ of $\bar{n}_{\rm [NII]}$ and $\bar{T}_{\rm [NII]}$. However, moving to the low-mass bin, the mean mass-weighted conditions shift such that the mass-weighted conditions lie just at the edge of $1\sigma$ from $\bar{n}_{\rm [NII]}$ and $\bar{T}_{\rm [NII]}$, although the $\bar{n}_{\rm [NII]}$ and $\bar{T}_{\rm [NII]}$ $1\sigma$ are well enclosed by the full $\bar{n}_{M}$ and $\bar{T}_{M}$ distributions.

The [\OIII] luminosity-weighted density and temperatures lie at or beyond $1\sigma$ range of both the \Halpha\, and [\NII] luminosity-weighted densities and temperatures. Across masses, the [\OIII] luminosity-weighted density and temperatures tend to be $\bar{n}_{\rm\,[OIII]} = 10^{-3.5 \pm 0.7} \rm cm^{-3}$ and $\bar{T}_{\rm [OIII]} = 10^{4.6 \pm 0.2} \rm K$, respectively. Further, $\bar{n}_{\rm\ [OIII]} $ lies systematically below $\bar{n}_M$ and $\bar{T}_{\rm\,[OIII]}$ above $\bar{T}_M$, indicating that [\OIII] emission traces preferentially more diffuse and warmer material. 

\textit{Metallicity:} We see little difference across the lines in the luminosity-weighted metallicities, with all three lines typically tracing $Z=10^{-0.9\pm0.2}Z_{\odot}$ and $Z=10^{-0.6\pm0.2}Z_{\odot}$ material for the low- and high-mass bins, respectively. Across lines, the luminosity-weighted metallicities and $1\sigma$ ranges are well enclosed by the full mass-weighted metallicity distribution, and the average mass-weighted metallicity lies within the $1\sigma$ ranges of luminosity-weighted values. All three trace preferentially higher-metallicity material than the mass-weighted metallicity does. This preference for metal-enriched material is expected for \NIIline\, and \OIIIline\, because their emissivities depend explicitly on gas metallicity. In contrast, we expect \Halpha\, also to be biased toward metal-rich material because metal cooling enhances the physical densities.

\subsection{Gas Kinematics}\label{disc:kinematic}

Figure \ref{fig:sb_kinesplit} shows the contribution from each kinematic phase to the SB maps in all three lines for a single disky galaxy ($M_\star = 10^{8.8} M_{\odot}$). We define inflows and outflows based on the virial velocity ($v_{\rm vir}$) of the halo following,
\begin{equation}
    v_{\rm vir} =\sqrt{G\, M_{\rm vir} / R_{\rm vir}}
\end{equation}
We then define inflowing particles as those with radial velocities ($v_r$) of $v_r < -0.1v_{\rm vir}$ and outflowing particles as $v_r > 0.1v_{\rm vir}$. These SB maps are generated by first selecting only inflowing or outflowing particles and then generating the projected SB map in each line.

We find inflows appear in emission as preferentially denser, filamentary structures. Outflows tend to be puffier and more volume-filling. Spatially, outflows tend to be biased along the semi-minor axis while inflows tend to be more circularly distributed. Importantly, there can be significant SB contributions from both kinematic phases in a given pixel.

\begin{figure*}
    \centering
    
    \includegraphics[width=0.98\linewidth]{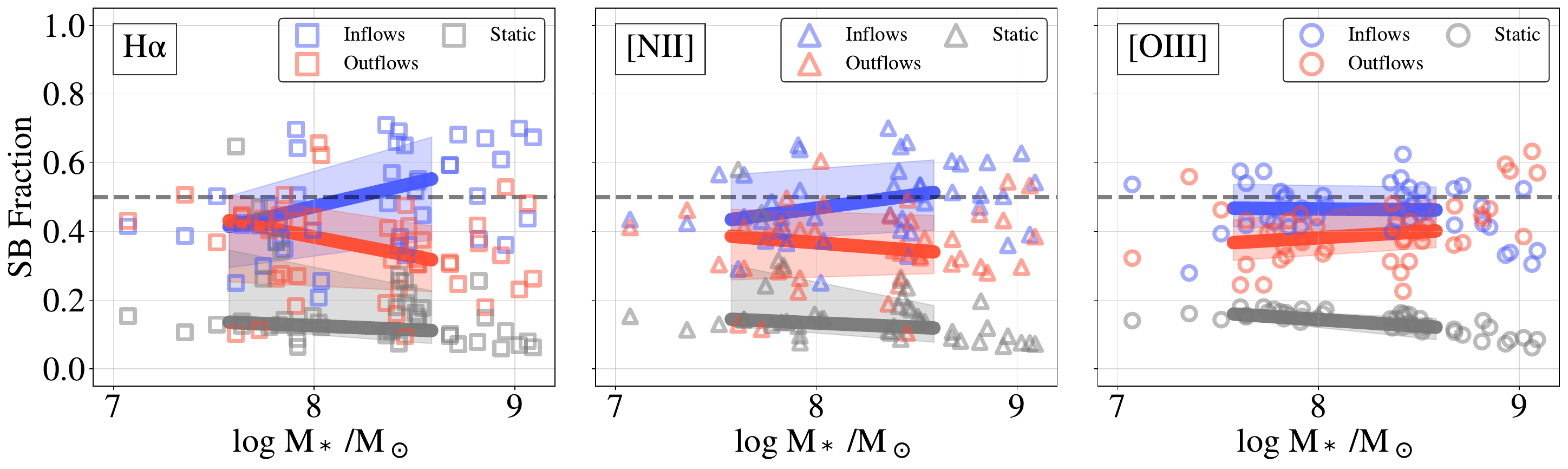}
    \caption{Fractional SB contribution from inflowing (blue), outflowing (red), and static gas (grey) as defined in Figure \ref{fig:sb_kinesplit}, to the total \HalpSB\, (left), \NIISB\, (center), and \OIIISB\, (right), with respect to galaxy stellar mass. We sum the total SB for all pixels within a radial annulus of ${\text \rHI}< b <0.5R_{\rm vir}$ in each map, then repeat this process for each kinematic phase to derive the SB fraction contributed by each. Solid lines represent the median SB fraction for each phase in the high- and low-mass bins, shaded bands show the $16-84\rm th$ percentile ranges.  While inflows contribute $\sim50\%$ of \HalpSB\, for high-mass galaxies, for other galaxy masses and the other emission lines, the total fractions tend to be within $0.2$ of one another. The similarity in total SB fractions suggests that the CGM of dwarf galaxies, as revealed by SB, is kinematically well mixed.}
    \label{fig:sb_kinematicfrac}
\end{figure*}

Within a given feedback feature, Figure \ref{fig:sb_kinesplit} reveals that features are kinematically complex and multiphase. For instance, for the SB feature in the outflowing maps in the northern portion of Figure \ref{fig:sb_kinesplit}, \Halpha\, and [\OIII] emission trace the puffy outflowing gas and shock front, while outflowing [\NII] emission tends to be clumpier in the same region. These variations across lines are in agreement with the luminosity-weighted conditions presented in Figure \ref{fig:conditons_summ}, which showed that [\NII] emission traces a narrow range in temperatures while \Halpha\, spans a wide range of temperatures, even reaching higher temperatures than [\NII]. For the inflowing component in that same region, we also find a shell of inflowing material surrounding the outflowing bubble primarily in the \OIIISB\, map and traces of it in the \HalpSB\, map.

Following the same procedure for SB covering fractions as in Figure \ref{fig:R_SB}, we find that the covering fractions for the two kinematic phases are within $20\%$ of each other for $80\%$ of our galaxies, across lines and SB thresholds. This similarity in extents of the two phases indicates that although inflows and outflows can appear morphologically dissimilar in these SB maps, they generally comprise similar total areas.


To fully quantify the contributions to SB from the two kinematic phases, we find the total SB from inflows, outflows, and static gas ($|v_r| < 0.1\,v_{\rm vir}$) by summing the total SB in a radial annulus of ${\text \rHI}< b<0.5R_{\rm vir}$ in each map (where $b$ is the projected impact parameter), and dividing it by the total SB map including all particles in the same annulus. We remove the galaxy from this calculation as we are most interested in discerning the SB signature from inflows and outflows in the CGM and disk-CGM interface. Figure \ref{fig:sb_kinematicfrac} shows the results for all galaxies included in this work. 

Across lines, inflows tend to contribute the largest fraction of SB. In particular, taking the median and $16-84\rm th$ percentile ranges across masses, inflows contribute $46_{-10} ^{+20\,}\%$, $49_{-10} ^{+12\,}\%$, and $47_{-5} ^{+7\,}\%$ of \Halpha, [\NII], and \OIIISB, respectively. On the other hand, outflows tend to contribute $36_{-14} ^{+9\,}\%$, $35_{-8} ^{+10\,}\%$, and $40_{-8} ^{+7\,}\%$ of \Halpha, [\NII], and \OIIISB, respectively. The SB fraction from static gas is typically $10-20\%$ and increases toward lower masses, which corresponds to the decrease in inflowing and outflowing SB fractions, so these phases contribute comparably, especially for \Halpha\ and [\NII].

Inflows contribute the dominant share of SB because they are preferentially denser and cooler than feedback-heated outflows and represent the largest fraction of gas mass. Using the same $v_r$ cut, inflows comprise $50\pm8\%$ of mass within $0.5\ {\text \Rvir}$ and outflows $33\pm7\%$ of mass, comparable to their respective SB fractions.

These results are similar to previous findings for simulations of higher-mass galaxies focusing on other metal lines. In FOGGIE, \citet{Lochhaas26} find that \OVI\, emission around MW-mass galaxies preferentially traces inflowing gas; however, outflows are critical in dispersing metals into the CGM inflows, and consequently the \OVI\, emission primarily traces the mixing of metal-enriched outflow remnants and inflowing gas accretion. This physical picture of emission tracing mixing material likely explains the shell of inflowing material shown in the \OIIISB\, map in Figure \ref{fig:sb_kinesplit}. However, for ${\text \Mstar}\sim 10^9 M_{\odot}$ galaxies in IllustrisTNG, outflows dominate the simulated \MgII\, luminosity fraction. Differences in subgrid physics could partly drive these differences; however, this also highlights the importance of using multiple lines to trace the CGM, since different lines trace different physical conditions.

\subsection{Correlations with Galaxy Properties}\label{disc:corr}


\begin{figure*}
    \centering
    
    \includegraphics[width=0.98\linewidth]{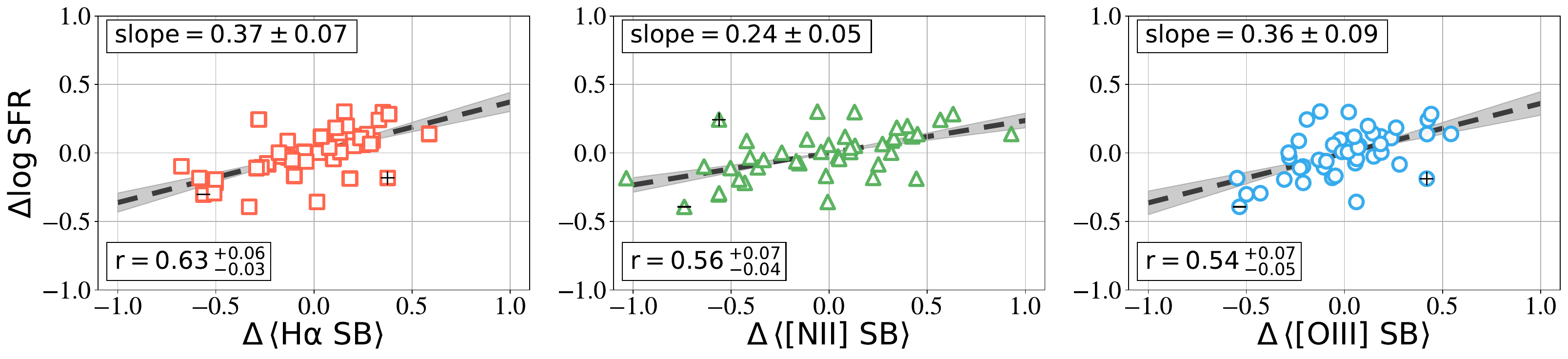}
    \caption{SB correlations with SFR. SB residuals (presented in Figure \ref{fig:sb_residuals}) are plotted on the horizontal axis, and SFR residuals ($\Delta \log \rm{SFR}$, derived from Figure \ref{fig:intro}) are plotted on the vertical axis. Linear regression fits and uncertainties are shown as grey dashed lines and shaded bands and slopes with uncertainties are reported in the top-left corner. Pearson correlation coefficients (r) are shown in the bottom-left corners. Correlations are computed for the full plotted sample, and the reported correlation ranges are computed via jackknifing. The range of minimum to maximum correlations from the jackknifing procedure is reported in the bottom-left corner. Across lines, we find significant ($r>0.5$) correlations between the SB and recent SFR. This correlation tends to be stronger for \HalpSB\, than \OIIISB\, and \NIISB. Generally, these correlations imply that the CGM tends to be brighter at a given mass for galaxies with enhanced SFRs. }
    \label{fig:SB_SFR}
\end{figure*}

\begin{table}
    \centering
    \begin{tabular}{c|ccc}
    \hline
     \textit{Slope} & \HalpSB & \NIISB & \OIIISB\\
          \hline
         SFR(10 Myr)& $0.40\pm0.08$ &$0.23\pm0.06$ &$0.35\pm0.10$  \\
        SFR(100 Myr)& ${0.37\pm0.07}$ & ${0.24\pm0.05}$ & ${0.36\pm0.09}$  \\
         SFR(500 Myr)& $0.35\pm0.06$ &$0.21\pm0.05$ &$0.33\pm0.07$  \\
          \hline\hline
          \textit{Correlation} & \HalpSB & \NIISB & \OIIISB\\
          \hline
         SFR(10 Myr)& $0.62^{+0.07}_{-0.02}$ & $0.49^{+0.14}_{-0.03}$ & $0.47^{+0.15}_{-0.02}$ \\
         SFR(100 Myr)& $0.63^{+0.06}_{-0.03}$ & $0.56^{+0.07}_{-0.04}$ & $0.54^{+0.07}_{-0.05}$ \\
         SFR(500 Myr)& $0.68^{+0.01}_{-0.08}$ & $0.58^{+0.05}_{-0.06}$ & $0.57^{+0.05}_{-0.08}$ \\
         \hline
    \end{tabular}
    \caption{Best-fit slopes (top) and Pearson correlation coefficients (bottom) between SB residuals and SFR residuals, with SFR averaged over $10$, $100$, and $500\,{\rm Myr}$. Values for $100\,{\rm Myr}$ are those shown in Figure \ref{fig:SB_SFR}. Slopes are generally steepest for the shortest SFR timescale, while correlations are generally strongest for the longest SFR timescale, although differences are small.}
    \label{tab:sfr_res}
\end{table}

In this section, we explore whether correlations exist between the magnitude of SB and galaxy properties. We explore correlations with star formation in Section \ref{corr:sfr} and galaxy size in Section \ref{corr:rHI}.

\subsubsection{Star Formation Rates}\label{corr:sfr}

Previous work has demonstrated a correlation between CGM properties and galaxy SFR. For instance, the CGM around star-forming MW-mass galaxies exhibits statistically elevated \OVI\, column densities compared to passive galaxies of the same mass \citep[e.g.,][]{Chen09,Tumlinson11,Tchernyshyov23}.  Here, we investigate if a similar correlation is discernible in the simulated SB.

Figure \ref{fig:SB_SFR} plots the SB residuals ($\Delta$) in each line (as in Figure \ref{fig:sb_residuals}) versus the SFR residuals measured over a $100\rm\, Myr$ timespan (derived from Figure \ref{fig:intro}). Linear best-fit lines and uncertainties are derived using \textsc{scipy} \citep{SciPy} and plotted in each panel. Exact values of fit slopes and their uncertainties are reported in the top-left corner of each panel. We also derive Pearson correlation coefficients using \textsc{scipy} and report them in the bottom-left corner. The quoted range on each correlation coefficient spans the minimum and maximum values obtained when performing a jackknifing procedure. The jackknifing procedure removes one galaxy, refits SB and SFR to derive residuals, and reports the correlation coefficient; this process is then repeated across the full sample. In each panel, a $+$ marks the galaxy whose removal most increases the correlation coefficient, and a $-$ marks the galaxy whose removal most decreases it. This selection of outliers is done for each line separately and thus the galaxies marked with a $+$ or $-$ are not necessarily the same galaxy across panels. 

The slopes of all three best-fit lines in Figure \ref{fig:SB_SFR} are positive.  Furthermore, the SFR and SB residuals have the same sign (i.e., enhanced or suppressed) for $82\%$, $73\%$, and $70\%$ of galaxies for \HalpSBres, \NIISBres, and \OIIISBres, respectively. Together, this indicates that enhanced SFR tends to correspond with enhanced SB.

\HalpSBres\, has the strongest correlations with $\Delta \log {\rm SFR}$ ($r=0.63$), while \NIISBres\, and \OIIISBres\, have weaker but still significant correlations of $r=0.56$ and $r=0.54$, respectively. The slopes of the best-fit line for \HalpSBres\, ($0.37\pm0.07$) and \OIIISBres\, ($0.36\pm0.09$) are the steepest. \NIISBres\, shows the shallowest slope of $0.24\pm 0.05$, which also is comparable to \HalpSBres. 

These correlation coefficients suggest that these lines have a dependence on galaxy activity, with \HalpSBres\ being the strongest correlation. These simulations do not explicitly track ionizing radiation from stars, and thus the correlations in Figure \ref{fig:SB_SFR} are more likely rooted in the physical conditions of the gas being shaped by stellar feedback. One explanation is that increased SFR, and therefore feedback, over the last $100\rm\ Myr$ drove heated, metal-enriched material from the disk into the CGM, increasing the current SB. In this case, [\OIII] emission traces the warmer components of the outflows while \Halpha\, emission traces the denser, $T\sim10^4 \rm\, K$ gas. [\NII] emission can still trace outflows (e.g., Figure \ref{fig:sb_kinesplit}); however, the lower ionization energy of \NII\, compared to \OIII\, limits the range of physical conditions where [\NII] emission is bright. Alternatively, periods of enhanced inflows could increase SB and, once they accrete onto the galaxy and form stars, could elevate the SFR. In this case, the accretion event would need to be sustained long enough to form the stars and still have accreting material in the CGM in order to appear in our SB maps. In other words, this scenario requires correlating past SB and accretion with current or future SFR, whereas thus far we have focused on past SFR and current SB. While this scenario is plausible and both this mechanism and its observable signatures will be explored in detail in future work, the correlations seen in  Figure \ref{fig:SB_SFR} are more likely a result of past outflows driving material into the CGM that appears in the current SB signatures.


\begin{figure*}
    \centering
    
    \includegraphics[width=0.98\linewidth]{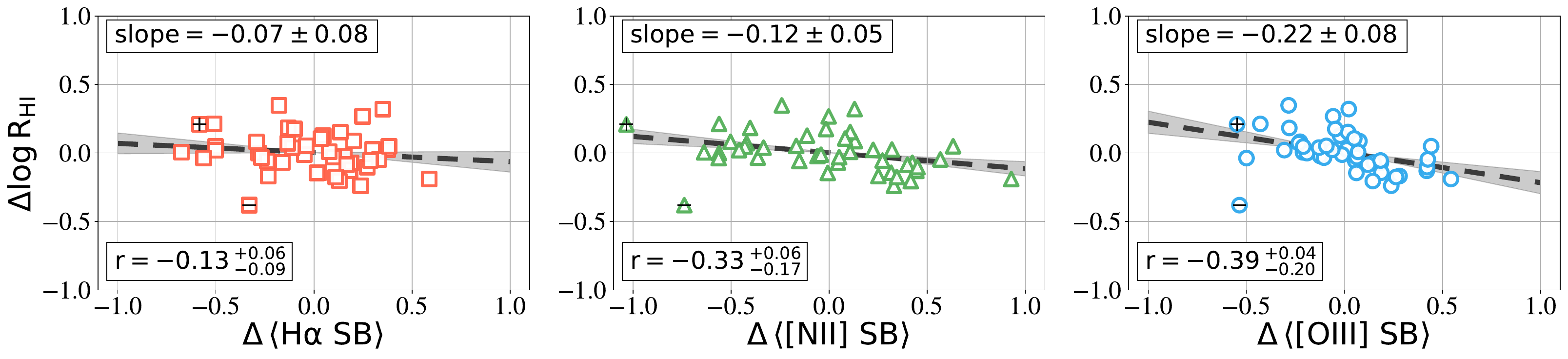}
    \caption{Identical to Figure \ref{fig:SB_SFR} but for \HI\, disk size (\rHI). \rHI\, residuals are computed from Figure \ref{fig:intro}. We find no significant correlation between \rHI\, and \HalpSB, but we find stronger ($r\approx -0.35$) anti-correlations for \NIISB\, and \OIIISB, implying that the more compact a galaxy is, the brighter the CGM tends to be in these metal lines. The range of errors from jackknifing tends to be larger for the \NIISB\, and \OIIISB\, correlations and is biased towards stronger (more negative) correlations.}
    \label{fig:SB_rhi}
\end{figure*}

Across lines, the jackknifing procedure yields a minimum correlation of $r=0.49$ for \OIIISB, suggesting the correlations reflect MMD rather than selection-criteria artifacts.
The range in correlation coefficients from the jackknifing procedure tends to be asymmetric and biased towards stronger correlations. For example, the lower bound for the \HalpSBres\, correlation is $r=0.60$ and an upper bound of $r=0.69$. The galaxy whose removal yields $r=0.69$ is marked with a $+$ and has an elevated SB due to a highly extended disk structure (and is also one of the low SB orientation angle outliers discussed in Section \ref{sec:sb_geom}). In other words, this galaxy has elevated \HalpSB\, driven more by its unique disk structure and formation than by recent SFR. The $+$ outlier for \NIISBres\, and \OIIISBres\, increases correlations by $0.07$. This outlier is the same in both \NIISBres\, and \OIIISBres\, and is a galaxy which recently underwent an extremely efficient feedback event that expelled a large portion of dense, metal-enriched material into the CGM and its SFR has since declined. This galaxy also appears in Figure \ref{fig:R_SB} as the most massive and extended galaxy in [\OIII] emission. While stellar feedback was the likely culprit in driving this mass ejection, it occurred more than $100\rm\, Myr$ in the past and thus does not affect the SFR residuals. These outliers show that unique galaxy formation and feedback events can produce unexpected behavior when linking galaxy activity to the SB, and they highlight the need for future studies to include a large sample of galaxies when mapping the baryon cycle. 


We find that the slopes and correlations persist when we change the timescale over which we measure SFR (reported in Table \ref{tab:sfr_res}). When measuring SFR over $10 ~\rm{Myr}$ and $500 ~\rm{Myr}$, best-fit line slopes fall well within the error ranges from the linear regression models, and correlation coefficients fall within the ranges from the jackknifing procedure. Although slopes and correlations vary slightly, slopes tend to be the shallowest for $500 ~\rm{Myr}$ and increase with shorter timescales. Correlation coefficients tend to generally increase with increasing timescales, suggesting that the present-day extended SB reflects star formation integrated over the past few hundred Myr rather than the instantaneous SFR. These results demonstrate that the correlations between galaxy activity and extended SB persist even when altering the timescales over which SFR is measured, suggesting that future observations should be able to use either \Halpha\, or UV imaging to measure SF and further test this theoretical prediction.

Previous theoretical work has also found correlations between gas emission and galaxy properties. Using IllustrisTNG, \citet{Nelson21} found that the \MgII\, luminosity correlates with the star formation surface density such that galaxies with enhanced \MgII\, luminosities at a fixed mass have more compact SF. However, when investigating a trend between specific SFR and \MgII\, luminosity, \citet{Nelson21} note no significant correlation. The ionization potential of \MgII\, ($7\rm\, eV$) is significantly lower than that of \NII\, ($14.5\rm\, eV$) and \OIII\, ($35\rm\, eV$), which may be partially responsible for the difference in SFR correlations. IllustrisTNG and MMD also model galactic winds differently, which may affect these correlations. In particular, IllustrisTNG models winds on a galactic scale with wind particles seeded with velocities, energy loadings, and mass loadings dependent on galaxy properties (\citealp[see][]{Pillepich18} for more details). Galactic winds in MMD are instead driven by SN events depositing energy into nearby gas particles, and outflows are subsequently driven by entropy \citep{Keller14}. Meanwhile, using the FOGGIE simulations, \citet{Lochhaas26} find that \OVI\, SB is enhanced around galaxies with ongoing SF events due to the influx of metals into the CGM, mirroring our results here. 

Observationally, a correlation between SFR and CGM emission has been shown in \citet{2026arXiv260718385Z}, which studied the \Halpha, [\OII], and [\OIII] emission around $10^{8.5}\leq {\text \Mstar} \leq 10^{11} M_{\odot}$ galaxies from the MaNGA survey using integral field spectroscopy. They find correlations between SFR and the SB measured at various factors of the galaxy's effective radius. They also find a difference in the slope of the emission radial profiles based on the SFR of the galaxy. Although we also find a correlation between SB and SFR, the authors of this observational study argue that such a relationship is due to the ionizing photons escaping into the CGM, which is not explicitly modeled in our suite.  Notably, these authors do not find a dependence on \Mstar\, whereas we find a clear mass dependence (Figure \ref{fig:Avg_SB}).

\subsubsection{HI Disk Size}\label{corr:rHI}

In addition to SFR correlations, we investigate correlations with the size of \HI\, disks (\rHI) in this section. We calculate \rHI\, residuals based on the best-fit line in Figure \ref{fig:intro} and then plot these \rHI\, residuals ($\Delta$\rHI) against the SB residuals in Figure \ref{fig:SB_rhi}. This plot is identical to Figure \ref{fig:SB_SFR} with best-fit slopes and correlation coefficients provided.

Unlike in Figure \ref{fig:SB_SFR}, all best-fit lines have negative slopes, and the SB and \rHI\, residuals have opposite signs for $60\%$, $64\%$, and $71\%$ of galaxies in \HalpSBres, \NIISBres, and \OIIISBres, respectively. Together, this indicates that more compact galaxy size tends to correspond with enhanced SB.

\HalpSB\, shows a non-significant correlation coefficient of $r=-0.13$ and correspondingly the flattest slope. By contrast, metal lines show more significant correlations. \NIISBres\, has a slope of $-0.12\pm0.05$ and correlation coefficient of $r=-0.33^{+0.06}_{-0.17}$ while \OIIISBres\, has the steepest slope of $-0.22\pm0.08$ and strongest correlation coefficient of $r=-0.39^{+0.04}_{-0.20}$. Notably, the jackknifing procedure shows that removing a single galaxy (marked by $-$) moves the correlations for \NIISBres\, and \OIIISBres\, to $-0.50$ and $-0.59$, respectively, suggesting a stronger correlation may be found with a larger sample of galaxies. This $-$ outlier galaxy (the same galaxy in all three lines) has a suppressed SB because it is gas-poor with $\sim1/4$ the median gas mass.

\begin{figure}
    \centering
    \includegraphics[width=0.97\linewidth]{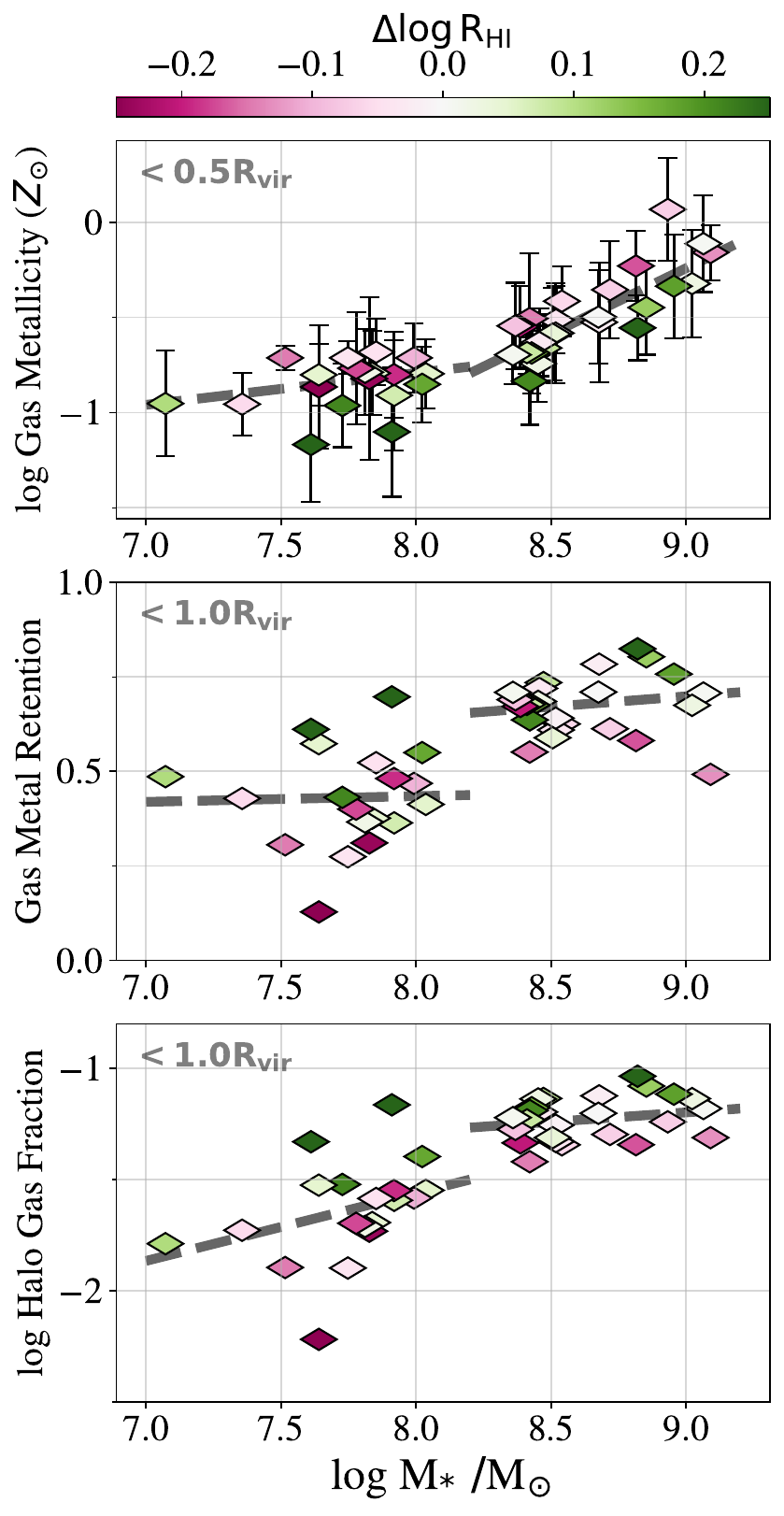}
    \caption{\textit{Top panel:} Median metallicity relative to solar ($Z/Z_{\odot}$) of all gas within $0.5 R_{\rm vir}$ versus stellar mass. All markers across panels are color-coded by \rHI\, residuals. Error bars show the standard deviation of particle data of the selected region. \textit{Middle panel:} Fraction of metals retained. This quantity is found by dividing the total oxygen mass for all gas mass within $R_{\rm vir}$ by the total oxygen mass produced over the galaxy's SFH, following \citet{Piacitelli26}. \textit{Bottom panel:} Total gas mass within $R_{\rm vir}$ divided by $M_{\rm vir}$. We find that compact galaxies ($\Delta{\text \rHI}<0$ pink markers) tend to have higher metallicity CGM, leading to the enhanced SB seen in Figure \ref{fig:SB_rhi}. These compact galaxies also have lower metal retention rates and lower gas fractions, indicating that these galaxies experience more efficient feedback that pushes metals and gas beyond the halo, compared to more extended galaxies. }
    \label{fig:rhi_Z}
\end{figure}

We further investigate the root cause of this relationship between SB and \rHI\, by plotting the median metallicity for all gas within $0.5R_{\rm vir}$ for each galaxy in our sample in Figure \ref{fig:rhi_Z}. Each galaxy is color-coded by its $\Delta$\rHI\, value, and error bars show the standard deviation of particle metallicity. As in previous plots, we plot best-fit lines as dashed lines. 

Compact galaxies ($\Delta {\text\rHI\,}<0$; $N=22$) in MMD tend to lie above the median line compared to more extended galaxies ($\Delta {\text\rHI\,}>0$; $N=22$). In particular, $77\%$ of compact galaxies lie above the median metallicity line and $82\%$ of extended galaxies lie below. Physically, this difference in metallicity is the key driver of the correlations in Figure \ref{fig:SB_rhi} as the gas metallicity directly affects the emissivity in our \cloudy\, modeling. These compact galaxies tend to have enhanced metallicities even when altering the volume of the region where metallicity is computed. More specifically, these compact galaxies have elevated metallicities compared to extended galaxies when considering regions: $R<{\text \rHI}$ (i.e., the galaxy disk), ${\text \rHI}<R< 0.5\ R_{\rm vir}$ (i.e., inner CGM), and $R<50\rm\, kpc$. Furthermore, we do not see any considerable correlations between  $\Delta {\text\rHI\,}$ and  $\Delta {\rm SFR}$ (Pearson correlation $r=0.1$), suggesting that the trends seen with \rHI\, have a different physical driver than the SFR trends. However, why galaxy size affects the metallicity of the inner halo is unclear.

The enhancements in gas metallicities imply that feedback is more efficient at dispersing metals from compact galaxies than from extended galaxies. If true, such a difference would also imprint on the evolution of the galaxy. Indeed, we see differences in terms of present-day morphology, where extended galaxies tend to have stronger stellar and gas disks with mean $\hat{z}$-direction angular momentum vectors of $j_{z,\star}^{\rm\  ext}=0.96$ and $j_{z,\rm gas}^{\rm\  ext}=0.88$ compared to the compact galaxies showing $j_{z,\star}^{\rm\ cmp}=0.80$ and $j_{z,\rm gas}^{\rm\ cmp}=0.68$. Although compact galaxies have lower angular momenta, $86\%$ of compact galaxies have stellar components that are kinematically classified as disks (following methods in Section \ref{sec:sim}).

The middle panel of Figure \ref{fig:rhi_Z} shows the gas metal retention. These values are computed by summing the total oxygen mass for all gas within $R_{\rm vir}$ ($M_{Z}$) and dividing it by the total oxygen mass formed by the galaxy over its star formation history. Total oxygen mass formed is computed following methods in \citet{Piacitelli26}, where an empirical metal yield $y$ is derived based on the total stellar and oxygen mass in the box. The total oxygen mass formed is then found with $M_{Z,\rm form} =  y \, {\text \Mstar}$ and metal retention is found by dividing the two masses: $M_{Z}/M_{Z,\rm form}$. We do not include metals locked in stars, which account for only $0.08\pm0.05$ of the metals produced by these galaxies \citep{Piacitelli25}. Including the stellar metal retention would raise each value in the middle panel of Figure~\ref{fig:rhi_Z} by roughly this amount. Because we compare compact and extended galaxies of similar stellar masses, this omission does not change our results. The bottom panel of Figure \ref{fig:rhi_Z} shows the halo gas fraction, defined by dividing the gas mass summed within $R_{\rm vir}$ by the virial mass.  

In metal retention, a separation between compact and extended galaxies reappears, with compact galaxies tending to show lower metal retention rates. However, the separations relative to the median are weaker than that seen in the metallicity plot. 
Specifically, $64\%$ of compact galaxies lie below the median and the same percentage of extended galaxies above. Similarly, in the bottom panel, we find that $77\%$ of compact galaxies lie below the median gas fraction line and $73\%$ of extended galaxies lie above. Although lower metal retention rates may initially seem contradictory to higher gas metallicities, the corresponding lower gas fractions in compact galaxies conspire to increase gas metallicity (i.e., higher metal mass fractions) and explain the trends across panels in Figure \ref{fig:rhi_Z}. Together, these results indicate that compact galaxies tend to have preferentially lost more metals and gas from their halos. As feedback is the primary driver of metal and energy transport from the disk, these results further indicate a difference in feedback efficiency between compact and extended galaxies. 


We emphasize that with only the $z=0$ snapshots included in this work, we cannot draw any conclusions on whether this increased feedback efficiency is what leads to more compact galaxies or if there is another variable that concurrently enables increased feedback efficiency and more compact galaxies. Given the differences in angular momentum between the two populations, the morphology could be important in determining or better understanding this feedback efficiency. In particular, compact galaxies likely have star formation that is more concentrated over their history, leading to more efficient feedback, compared to extended galaxies with star formation spread over the disk.  Answering these questions will require expanding the analysis to the full evolution of these galaxies and will be explored in future work. Nonetheless, the results we present show how differences in evolution can manifest as discernible present-day signatures in the extended SB of dwarf galaxies. 

\section{Discussion}\label{sec:disc}
\subsection{Limitations}

As stated in Section \ref{sec:sim}, we do not model line-of-sight radiative transfer when creating the SB maps; that is, we assume that no absorption or scattering occurs after the photon is emitted. We make this simplifying assumption to reduce modeling uncertainties and to address the current lack of theoretical work studying optical CGM emission from dwarf galaxies.  Attenuation is likely impactful for the \Halpha\, emission, especially within the ISM where densities are the highest. By assuming the medium is optically thin, our median SB results for each galaxy are upper limits, and the SB at small projected distances for a real galaxy is likely fainter. For this reason, the correlations we present in Section \ref{sec:drivers} between SB and galaxy properties may be obscured in observations affected by radiative transfer. Although the ISM SB is likely overestimated, this paper primarily focuses on the extended SB from the inner CGM of dwarfs, where gas densities are much lower.

Since emission scales as $n^2$, a central limitation of our study is the numerical resolution of our simulations. Previous work from the FOGGIE collaboration found that increased spatial resolution for the $z=3$ CGM of a MW-mass galaxy can impact the total luminosity and the distribution in various lines \citep{Corlies20}, e.g., increasing the luminosity of \Halpha\, by a factor of two when increasing the spatial resolution in the CGM from $2200\rm\, pc$ to $68\rm\, pc$. Although not directly comparable, the SPH smoothing length for MMD within the inner CGM is $\sim1.5\pm0.8 \rm\, kpc$. Increasing the mass resolution of these simulations could therefore result in higher luminosities than those presented here, particularly for \Halpha\, and [\NII] which trace denser phases (Figure \ref{fig:conditons_summ}). Additionally, increased resolution would produce smaller cloud sizes and more substructure in the resulting SB maps.



\section{Summary}\label{sec:summary}
We present an analysis of simulated SB in three optical lines which will be surveyed by the upcoming MOTHRA telephoto array: \Halpha, \NIIline, and \OIIIline. This work is the first in a series that intends to develop a useful theoretical framework for interpreting the CGM SB around dwarf galaxies. Of particular interest is the connection between CGM SB and galaxy properties, as well as how CGM SB may trace or reveal the baryon cycle of dwarfs. Our main findings are as follows:

\begin{enumerate}

    \item Median SB values in all three lines decline with declining stellar mass, although metal lines decline much more steeply than \HalpSB\, due to declining metallicities (Figure \ref{fig:Avg_SB}). Median SB values at a given galaxy mass can vary significantly ($\pm 1 \rm\, dex$) with important driving factors from the galaxy activity and size. 
    \item Enhanced or suppressed SB across lines correlate well with one another, indicating that enhancements in CGM SB in one line typically coincide with enhancements in the SB in other lines. Furthermore, 
    \begin{enumerate}
        \item SB residuals tend to positively correlate with SFR over a range of timescales and with correlation coefficients of $r\sim 0.5$ or higher (Figure \ref{fig:SB_SFR}). This correlation demonstrates the galaxy-CGM connection suggested in previous work and indicates that CGM SB mapping may hold strong promise in mapping the effects of feedback on the CGM. However, our simulations suggest a stronger correlation between SB and SFR measured over longer timescales, motivating UV-continuum SFR measurements to test this correlation observationally.

        \item SB residuals tend to negatively correlate with \HI\, disk size. However, compared to the correlations with SFR, the trend with \HI\, disk size is weaker (Figure \ref{fig:SB_rhi}). 
        If true, this trend implies that galaxies with more extended \HI\, disks have suppressed SB in metal lines compared to more compact dwarf galaxies. This trend would be driven by a difference in how effectively feedback-driven material exits the disk of the galaxy and enriches the CGM. Indeed, we find galaxies with enhanced gas-phase metallicities and lower metal retention rates and gas fractions tend to have less extended \HI\, disks (Figure \ref{fig:rhi_Z}). Observationally, resolved 21-cm imaging of dwarf galaxies, paired with deep [\NII] and [\OIII] SB maps, could test whether metal-line SB is suppressed in galaxies with more extended \HI\ disks at fixed stellar mass.
        \end{enumerate}
    \item SB in all three lines can cover significant fractions of the halo size, especially at the higher masses (Figure \ref{fig:R_SB}). In particular,
    \begin{enumerate}
        \item Regions with $\log \rm H\alpha\, SB > -20$ and $> -22$ tend to cover a consistent $\sim5\%$ and $\sim30\%$ of the halo across stellar masses, suggesting that \Halpha\, emission will be useful in tracing the halo gas of dwarf galaxies.
        \item The extent of [\NII] and [\OIII] emission tends to have a stronger mass dependence than \Halpha\, due to their explicit metallicity dependence. 
    \end{enumerate}

    \item Contours at the same SB limits as above tend to be highly eccentric and oriented preferentially along the semi-minor axis (Figure \ref{fig:SB_shape}), indicating that CGM SB for dwarfs may be significantly sourced by outflows.

\end{enumerate}

Mapping the CGM in emission is a critical next step in understanding the CGM and the baryon cycle of dwarf galaxies. This work indicates that there is a connection between galaxy properties and CGM SB, and observations from MOTHRA as well as KCWI and MUSE will be critical in constraining this further. With a large sample of observed galaxies and joint \HI\, and UV observations, observations can test the correlations seen in Figures \ref{fig:SB_SFR} and \ref{fig:SB_rhi}. Additionally, we provide total luminosities of each line in Figure \ref{fig:Avg_SB}, which can serve as useful new observables to constrain simulations. Finally, the well-mixed CGM properties in kinematic space as revealed by SB indicate that we must be cautious when discerning between inflowing and outflowing material in SB maps of real galaxies. This work is not an exhaustive study of how to identify inflows and outflows in SB maps, and future work will aim to investigate if discernible trends exist in kinematic and SB space.

\begin{acknowledgments}

We thank Cameron Hummels for helpful conversations during this work. Resources supporting this work were provided by the NASA High-End Computing (HEC) Program through the NASA Advanced Supercomputing (NAS) Division at Ames Research Center. Some of the simulations were performed using resources made available by the Flatiron Institute. The Flatiron Institute is a division of the Simons Foundation. This work used Stampede2 at the Texas Advanced Computing Center (TACC) through allocation MCA94P018 from the Advanced Cyberinfrastructure Coordination Ecosystem: Services \& Support (ACCESS) program, which is supported by U.S. National Science Foundation grants \#2138259, \#2138286, \#2138307, \#2137603, and \#2138296. D.R.P and A.M.B are supported by NASA Grant 80NSSC24K0894. A.M.B acknowledges support by grant FI-CCA-Research-00011826 from the Simons Foundation. J.W. is supported by a grant from NSERC (National Science and Engineering Research Council) Canada. C.C. was supported by the NSF under CAREER grant AST-1848107, and this work was performed in part at Aspen Center for Physics, which is supported by NSF grant PHY-2210452. N.N.S. was supported by the National Science Foundation MPS-Ascend award ID 2212959. B.W.K. acknowledges support provided by NASA through a grant from the Space Telescope Science Institute, through grant HST AR-17547.

\end{acknowledgments}




\bibliography{main}{}

@ARTICLE{HM2012,
       author = {{Haardt}, Francesco and {Madau}, Piero},
        title = "{Radiative Transfer in a Clumpy Universe. IV. New Synthesis Models of the Cosmic UV/X-Ray Background}",
      journal = {\apj},
         year = 2012,
        month = feb,
       volume = {746},
       number = {2},
          eid = {125},
        pages = {125},
          doi = {10.1088/0004-637X/746/2/125},
       adsurl = {https://ui.adsabs.harvard.edu/abs/2012ApJ...746..125H}
}

@ARTICLE{Kroupa2001_IMF,
       author = {{Kroupa}, Pavel},
        title = "{On the variation of the initial mass function}",
      journal = {\mnras},
         year = 2001,
        month = apr,
       volume = {322},
       number = {2},
        pages = {231-246},
          doi = {10.1046/j.1365-8711.2001.04022.x},
archivePrefix = {arXiv},
       eprint = {astro-ph/0009005},
 primaryClass = {astro-ph},
       adsurl = {https://ui.adsabs.harvard.edu/abs/2001MNRAS.322..231K}
}

@ARTICLE{CHANGA,
       author = {{Menon}, Harshitha and {Wesolowski}, Lukasz and {Zheng}, Gengbin and {Jetley}, Pritish and {Kale}, Laxmikant and {Quinn}, Thomas and {Governato}, Fabio},
        title = "{Adaptive techniques for clustered N-body cosmological simulations}",
      journal = {Computational Astrophysics and Cosmology},
         year = 2015,
        month = mar,
       volume = {2},
          eid = {1},
        pages = {1},
          doi = {10.1186/s40668-015-0007-9},
archivePrefix = {arXiv},
       eprint = {1409.1929},
 primaryClass = {astro-ph.IM},
       adsurl = {https://ui.adsabs.harvard.edu/abs/2015ComAC...2....1M}
}

@ARTICLE{Shen2010,
       author = {{Shen}, S. and {Wadsley}, J. and {Stinson}, G.},
        title = "{The enrichment of the intergalactic medium with adiabatic feedback - I. Metal cooling and metal diffusion}",
      journal = {\mnras},
         year = 2010,
        month = sep,
       volume = {407},
       number = {3},
        pages = {1581-1596},
          doi = {10.1111/j.1365-2966.2010.17047.x},
archivePrefix = {arXiv},
       eprint = {0910.5956},
 primaryClass = {astro-ph.CO},
       adsurl = {https://ui.adsabs.harvard.edu/abs/2010MNRAS.407.1581S}
}

@ARTICLE{Christensen2012,
       author = {{Christensen}, Charlotte and {Quinn}, Thomas and {Governato}, Fabio and {Stilp}, Adrienne and {Shen}, Sijing and {Wadsley}, James},
        title = "{Implementing molecular hydrogen in hydrodynamic simulations of galaxy formation}",
      journal = {\mnras},
         year = 2012,
        month = oct,
       volume = {425},
       number = {4},
        pages = {3058-3076},
          doi = {10.1111/j.1365-2966.2012.21628.x},
archivePrefix = {arXiv},
       eprint = {1205.5567},
 primaryClass = {astro-ph.CO},
       adsurl = {https://ui.adsabs.harvard.edu/abs/2012MNRAS.425.3058C}
}

@ARTICLE{Ruan2025,
       author = {{Ruan}, Dilys and {Brooks}, Alyson M. and {Cruz}, Akaxia and {Peter}, Annika H.~G. and {Keller}, Benjamin W. and {Quinn}, Thomas and {Wadsley}, James and {Adams}, Elizabeth A.~K.},
        title = "{Predictions for detecting a turndown in the baryonic Tully{\textendash}Fisher relation}",
      journal = {\mnras},
         year = 2025,
        month = aug,
       volume = {541},
       number = {3},
        pages = {2180-2196},
          doi = {10.1093/mnras/staf1099},
archivePrefix = {arXiv},
       eprint = {2503.16607},
 primaryClass = {astro-ph.GA},
       adsurl = {https://ui.adsabs.harvard.edu/abs/2025MNRAS.541.2180R}
}

@ARTICLE{Keith2025,
       author = {{Keith}, Blake and {Munshi}, Ferah and {Brooks}, Alyson M. and {Van Nest}, Jordan and {Engelhardt}, Anna and {Cruz}, Akaxia and {Keller}, Ben and {Quinn}, Thomas and {Wadsley}, James},
        title = "{A MARVEL-ous Study of How Well Galaxy Shapes Reflect Dark Matter Halo Shapes in Cold Dark Matter Simulations}",
      journal = {\apj},
         year = 2025,
        month = jun,
       volume = {986},
       number = {2},
          eid = {138},
        pages = {138},
          doi = {10.3847/1538-4357/add40d},
archivePrefix = {arXiv},
       eprint = {2501.16317},
 primaryClass = {astro-ph.GA},
       adsurl = {https://ui.adsabs.harvard.edu/abs/2025ApJ...986..138K}
}

@ARTICLE{Piacitelli25,
       author = {{Piacitelli}, Daniel R. and {Brooks}, Alyson M. and {Christensen}, Charlotte and {Sanchez}, N. Nicole and {Faerman}, Yakov and {Shen}, Sijing and {Cruz}, Akaxia and {Keller}, Ben and {Quinn}, Thomas R. and {Wadsley}, James},
        title = "{Marvelous Metals: Surveying the Circumgalactic Medium of Simulated Dwarf Galaxies}",
      journal = {\apj},
         year = 2025,
        month = nov,
       volume = {993},
       number = {2},
          eid = {230},
        pages = {230},
          doi = {10.3847/1538-4357/ae06a0},
archivePrefix = {arXiv},
       eprint = {2505.08861},
 primaryClass = {astro-ph.GA},
       adsurl = {https://ui.adsabs.harvard.edu/abs/2025ApJ...993..230P}
}

@ARTICLE{Bordoloi_14,
       author = {{Bordoloi}, Rongmon and {Tumlinson}, Jason and {Werk}, Jessica K. and {Oppenheimer}, Benjamin D. and {Peeples}, Molly S. and {Prochaska}, J. Xavier and {Tripp}, Todd M. and {Katz}, Neal and {Dav{\'e}}, Romeel and {Fox}, Andrew J. and {Thom}, Christopher and {Ford}, Amanda Brady and {Weinberg}, David H. and {Burchett}, Joseph N. and {Kollmeier}, Juna A.},
        title = "{The COS-Dwarfs Survey: The Carbon Reservoir around Sub-L* Galaxies}",
      journal = {\apj},
         year = 2014,
        month = dec,
       volume = {796},
       number = {2},
          eid = {136},
        pages = {136},
          doi = {10.1088/0004-637X/796/2/136},
archivePrefix = {arXiv},
       eprint = {1406.0509},
 primaryClass = {astro-ph.GA},
       adsurl = {https://ui.adsabs.harvard.edu/abs/2014ApJ...796..136B}
}

@ARTICLE{Johnson_17,
       author = {{Johnson}, Sean D. and {Chen}, Hsiao-Wen and {Mulchaey}, John S. and {Schaye}, Joop and {Straka}, Lorrie A.},
        title = "{The Extent of Chemically Enriched Gas around Star-forming Dwarf Galaxies}",
      journal = {\apjl},
         year = 2017,
        month = nov,
       volume = {850},
       number = {1},
          eid = {L10},
        pages = {L10},
          doi = {10.3847/2041-8213/aa9370},
archivePrefix = {arXiv},
       eprint = {1710.06441},
 primaryClass = {astro-ph.GA},
       adsurl = {https://ui.adsabs.harvard.edu/abs/2017ApJ...850L..10J}
}

@ARTICLE{Mishra_24,
       author = {{Mishra}, Nishant and {Johnson}, Sean D. and {Rudie}, Gwen C. and {Chen}, Hsiao-Wen and {Schaye}, Joop and {Qu}, Zhijie and {Zahedy}, Fakhri S. and {Boettcher}, Erin T. and {Cantalupo}, Sebastiano and {Chen}, Mandy C. and {Faucher-Gigu{\'e}re}, Claude-Andr{\'e} and {Greene}, Jenny E. and {Li}, Jennifer I. -Hsiu and {Liu}, Zhuoqi (Will) and {Lopez}, Sebastian and {Petitjean}, Patrick},
        title = "{The Cosmic Ultraviolet Baryon Survey (CUBS). IX. The Enriched Circumgalactic and Intergalactic Medium Around Star-forming Field Dwarf Galaxies Traced by O VI Absorption}",
      journal = {\apj},
         year = 2024,
        month = nov,
       volume = {976},
       number = {1},
          eid = {149},
        pages = {149},
          doi = {10.3847/1538-4357/ad7b0a},
archivePrefix = {arXiv},
       eprint = {2408.11151},
 primaryClass = {astro-ph.GA},
       adsurl = {https://ui.adsabs.harvard.edu/abs/2024ApJ...976..149M}
}

@ARTICLE{Dutta_25_col,
       author = {{Dutta}, Sayak and {Muzahid}, Sowgat and {Schaye}, Joop and {Bouch{\'e}}, Nicolas F. and {Cantalupo}, Sebastiano and {Chen}, Hsiao-Wen and {Johnson}, Sean},
        title = "{MUSEQuBES: The Column Density, Covering Fraction, and Mass of O VI-bearing Gas in and Around Low-redshift Galaxies}",
      journal = {\apj},
         year = 2025,
        month = may,
       volume = {985},
       number = {1},
          eid = {44},
        pages = {44},
          doi = {10.3847/1538-4357/adc922},
archivePrefix = {arXiv},
       eprint = {2409.15423},
 primaryClass = {astro-ph.GA},
       adsurl = {https://ui.adsabs.harvard.edu/abs/2025ApJ...985...44D}
}

@ARTICLE{Tchernyshyov_22,
       author = {{Tchernyshyov}, Kirill and {Werk}, Jessica K. and {Wilde}, Matthew C. and {Prochaska}, J. Xavier and {Tripp}, Todd M. and {Burchett}, Joseph N. and {Bordoloi}, Rongmon and {Howk}, J. Christopher and {Lehner}, Nicolas and {O'Meara}, John M. and {Tejos}, Nicolas and {Tumlinson}, Jason},
        title = "{The CGM$^{2}$ Survey: Circumgalactic O VI from Dwarf to Massive Star-forming Galaxies}",
      journal = {\apj},
         year = 2022,
        month = mar,
       volume = {927},
       number = {2},
          eid = {147},
        pages = {147},
          doi = {10.3847/1538-4357/ac450c},
archivePrefix = {arXiv},
       eprint = {2110.13167},
 primaryClass = {astro-ph.GA},
       adsurl = {https://ui.adsabs.harvard.edu/abs/2022ApJ...927..147T}
}

@ARTICLE{Tremmel17,
       author = {{Tremmel}, M. and {Karcher}, M. and {Governato}, F. and {Volonteri}, M. and {Quinn}, T.~R. and {Pontzen}, A. and {Anderson}, L. and {Bellovary}, J.},
        title = "{The Romulus cosmological simulations: a physical approach to the formation, dynamics and accretion models of SMBHs}",
      journal = {\mnras},
         year = 2017,
        month = sep,
       volume = {470},
       number = {1},
        pages = {1121-1139},
          doi = {10.1093/mnras/stx1160},
archivePrefix = {arXiv},
       eprint = {1607.02151},
 primaryClass = {astro-ph.GA},
       adsurl = {https://ui.adsabs.harvard.edu/abs/2017MNRAS.470.1121T}
}

@ARTICLE{Planck16cosmo,
       author = {{Planck Collaboration} and {Ade}, P.~A.~R. and {Aghanim}, N. and {Arnaud}, M. and {Ashdown}, M. and {Aumont}, J. and {Baccigalupi}, C. and {Banday}, A.~J. and {Barreiro}, R.~B. and {Bartlett}, J.~G. and {Bartolo}, N. and {Battaner}, E. and {Battye}, R. and {Benabed}, K. and {Beno{\^\i}t}, A. and {Benoit-L{\'e}vy}, A. and {Bernard}, J.-P. and {Bersanelli}, M. and {Bielewicz}, P. and {Bock}, J.~J. and {Bonaldi}, A. and {Bonavera}, L. and {Bond}, J.~R. and {Borrill}, J. and {Bouchet}, F.~R. and {Boulanger}, F. and {Bucher}, M. and {Burigana}, C. and {Butler}, R.~C. and {Calabrese}, E. and {Cardoso}, J.-F. and {Catalano}, A. and {Challinor}, A. and {Chamballu}, A. and {Chary}, R.-R. and {Chiang}, H.~C. and {Chluba}, J. and {Christensen}, P.~R. and {Church}, S. and {Clements}, D.~L. and {Colombi}, S. and {Colombo}, L.~P.~L. and {Combet}, C. and {Coulais}, A. and {Crill}, B.~P. and {Curto}, A. and {Cuttaia}, F. and {Danese}, L. and {Davies}, R.~D. and {Davis}, R.~J. and {de Bernardis}, P. and {de Rosa}, A. and {de Zotti}, G. and {Delabrouille}, J. and {D{\'e}sert}, F.-X. and {Di Valentino}, E. and {Dickinson}, C. and {Diego}, J.~M. and {Dolag}, K. and {Dole}, H. and {Donzelli}, S. and {Dor{\'e}}, O. and {Douspis}, M. and {Ducout}, A. and {Dunkley}, J. and {Dupac}, X. and {Efstathiou}, G. and {Elsner}, F. and {En{\ss}lin}, T.~A. and {Eriksen}, H.~K. and {Farhang}, M. and {Fergusson}, J. and {Finelli}, F. and {Forni}, O. and {Frailis}, M. and {Fraisse}, A.~A. and {Franceschi}, E. and {Frejsel}, A. and {Galeotta}, S. and {Galli}, S. and {Ganga}, K. and {Gauthier}, C. and {Gerbino}, M. and {Ghosh}, T. and {Giard}, M. and {Giraud-H{\'e}raud}, Y. and {Giusarma}, E. and {Gjerl{\o}w}, E. and {Gonz{\'a}lez-Nuevo}, J. and {G{\'o}rski}, K.~M. and {Gratton}, S. and {Gregorio}, A. and {Gruppuso}, A. and {Gudmundsson}, J.~E. and {Hamann}, J. and {Hansen}, F.~K. and {Hanson}, D. and {Harrison}, D.~L. and {Helou}, G. and {Henrot-Versill{\'e}}, S. and {Hern{\'a}ndez-Monteagudo}, C. and {Herranz}, D. and {Hildebrandt}, S.~R. and {Hivon}, E. and {Hobson}, M. and {Holmes}, W.~A. and {Hornstrup}, A. and {Hovest}, W. and {Huang}, Z. and {Huffenberger}, K.~M. and {Hurier}, G. and {Jaffe}, A.~H. and {Jaffe}, T.~R. and {Jones}, W.~C. and {Juvela}, M. and {Keih{\"a}nen}, E. and {Keskitalo}, R. and {Kisner}, T.~S. and {Kneissl}, R. and {Knoche}, J. and {Knox}, L. and {Kunz}, M. and {Kurki-Suonio}, H. and {Lagache}, G. and {L{\"a}hteenm{\"a}ki}, A. and {Lamarre}, J.-M. and {Lasenby}, A. and {Lattanzi}, M. and {Lawrence}, C.~R. and {Leahy}, J.~P. and {Leonardi}, R. and {Lesgourgues}, J. and {Levrier}, F. and {Lewis}, A. and {Liguori}, M. and {Lilje}, P.~B. and {Linden-V{\o}rnle}, M. and {L{\'o}pez-Caniego}, M. and {Lubin}, P.~M. and {Mac{\'\i}as-P{\'e}rez}, J.~F. and {Maggio}, G. and {Maino}, D. and {Mandolesi}, N. and {Mangilli}, A. and {Marchini}, A. and {Maris}, M. and {Martin}, P.~G. and {Martinelli}, M. and {Mart{\'\i}nez-Gonz{\'a}lez}, E. and {Masi}, S. and {Matarrese}, S. and {McGehee}, P. and {Meinhold}, P.~R. and {Melchiorri}, A. and {Melin}, J.-B. and {Mendes}, L. and {Mennella}, A. and {Migliaccio}, M. and {Millea}, M. and {Mitra}, S. and {Miville-Desch{\^e}nes}, M.-A. and {Moneti}, A. and {Montier}, L. and {Morgante}, G. and {Mortlock}, D. and {Moss}, A. and {Munshi}, D. and {Murphy}, J.~A. and {Naselsky}, P. and {Nati}, F. and {Natoli}, P. and {Netterfield}, C.~B. and {N{\o}rgaard-Nielsen}, H.~U. and {Noviello}, F. and {Novikov}, D. and {Novikov}, I. and {Oxborrow}, C.~A. and {Paci}, F. and {Pagano}, L. and {Pajot}, F. and {Paladini}, R. and {Paoletti}, D. and {Partridge}, B. and {Pasian}, F. and {Patanchon}, G. and {Pearson}, T.~J. and {Perdereau}, O. and {Perotto}, L. and {Perrotta}, F. and {Pettorino}, V. and {Piacentini}, F. and {Piat}, M. and {Pierpaoli}, E. and {Pietrobon}, D. and {Plaszczynski}, S. and {Pointecouteau}, E. and {Polenta}, G. and {Popa}, L. and {Pratt}, G.~W. and {Pr{\'e}zeau}, G.},
        title = "{Planck 2015 results. XIII. Cosmological parameters}",
      journal = {\aap},
         year = 2016,
        month = sep,
       volume = {594},
          eid = {A13},
        pages = {A13},
          doi = {10.1051/0004-6361/201525830},
archivePrefix = {arXiv},
       eprint = {1502.01589},
 primaryClass = {astro-ph.CO},
       adsurl = {https://ui.adsabs.harvard.edu/abs/2016A&A...594A..13P}
}

@ARTICLE{GenetIC2021,
       author = {{Stopyra}, Stephen and {Pontzen}, Andrew and {Peiris}, Hiranya and {Roth}, Nina and {Rey}, Martin P.},
        title = "{GenetIC{\textemdash}A New Initial Conditions Generator to Support Genetically Modified Zoom Simulations}",
      journal = {\apjs},
         year = 2021,
        month = feb,
       volume = {252},
       number = {2},
          eid = {28},
        pages = {28},
          doi = {10.3847/1538-4365/abcd94},
archivePrefix = {arXiv},
       eprint = {2006.01841},
 primaryClass = {astro-ph.IM},
       adsurl = {https://ui.adsabs.harvard.edu/abs/2021ApJS..252...28S}
}

@ARTICLE{Keller14,
       author = {{Keller}, B.~W. and {Wadsley}, J. and {Benincasa}, S.~M. and {Couchman}, H.~M.~P.},
        title = "{A superbubble feedback model for galaxy simulations}",
      journal = {\mnras},
         year = 2014,
        month = aug,
       volume = {442},
       number = {4},
        pages = {3013-3025},
          doi = {10.1093/mnras/stu1058},
archivePrefix = {arXiv},
       eprint = {1405.2625},
 primaryClass = {astro-ph.GA},
       adsurl = {https://ui.adsabs.harvard.edu/abs/2014MNRAS.442.3013K}
}

@ARTICLE{Asplund2009,
       author = {{Asplund}, Martin and {Grevesse}, Nicolas and {Sauval}, A. Jacques and {Scott}, Pat},
        title = "{The Chemical Composition of the Sun}",
      journal = {\araa},
         year = 2009,
        month = sep,
       volume = {47},
       number = {1},
        pages = {481-522},
          doi = {10.1146/annurev.astro.46.060407.145222},
archivePrefix = {arXiv},
       eprint = {0909.0948},
 primaryClass = {astro-ph.SR},
       adsurl = {https://ui.adsabs.harvard.edu/abs/2009ARA&A..47..481A}
}

@misc{pynbody,
  author = {{Pontzen}, A. and {Ro{\v s}kar}, R. and {Stinson}, G.~S. and {Woods},
     R. and {Reed}, D.~M. and {Coles}, J. and {Quinn}, T.~R.},
  title = "{pynbody: Astrophysics Simulation Analysis for Python}",
  note = {Astrophysics Source Code Library, ascl:1305.002},
  year = 2013
}

@ARTICLE{AHF_09,
       author = {{Knollmann}, Steffen R. and {Knebe}, Alexander},
        title = "{AHF: Amiga's Halo Finder}",
      journal = {\apjs},
         year = 2009,
        month = jun,
       volume = {182},
       number = {2},
        pages = {608-624},
          doi = {10.1088/0067-0049/182/2/608},
archivePrefix = {arXiv},
       eprint = {0904.3662},
 primaryClass = {astro-ph.CO},
       adsurl = {https://ui.adsabs.harvard.edu/abs/2009ApJS..182..608K}
}

@ARTICLE{Cruz_25,
       author = {{Cruz}, Akaxia and {Brooks}, Alyson and {Lisanti}, Mariangela and {Peter}, Annika H.~G. and {Geda}, Robel and {Quinn}, Thomas and {Tremmel}, Michael and {Munshi}, Ferah and {Keller}, Ben and {Wadsley}, James},
        title = "{Dwarf diversity in $Λ$CDM with baryons}",
      journal = {arXiv e-prints},
         year = 2025,
        month = oct,
          eid = {arXiv:2510.11800},
        pages = {arXiv:2510.11800},
          doi = {10.48550/arXiv.2510.11800},
archivePrefix = {arXiv},
       eprint = {2510.11800},
 primaryClass = {astro-ph.GA},
       adsurl = {https://ui.adsabs.harvard.edu/abs/2025arXiv251011800C}
}

@ARTICLE{Cloudy_Ferland17,
   author = {{Ferland}, G.~J. and {Chatzikos}, M. and {Guzm{\'a}n}, F. and {Lykins}, M.~L. and {van Hoof}, P.~A.~M. and {Williams}, R.~J.~R. and {Abel}, N.~P. and {Badnell}, N.~R. and {Keenan}, F.~P. and {Porter}, R.~L. and {Stancil}, P.~C.},
    title = "{The 2017 Release Cloudy}",
  journal = {\rmxaa},
archivePrefix = "arXiv",
   eprint = {1705.10877},
     year = 2017,
    month = oct,
   volume = 53,
    pages = {385-438},
   adsurl = {http://adsabs.harvard.edu/abs/2017RMxAA..53..385F}}

@ARTICLE{Bland_Tully88,
       author = {{Bland}, Jonathan and {Tully}, Brent},
        title = "{Large-scale bipolar wind in M82}",
      journal = {\nat},
         year = 1988,
        month = jul,
       volume = {334},
       number = {6177},
        pages = {43-45},
          doi = {10.1038/334043a0},
       adsurl = {https://ui.adsabs.harvard.edu/abs/1988Natur.334...43B}
}

@ARTICLE{Keres05,
       author = {{Kere{\v{s}}}, Du{\v{s}}an and {Katz}, Neal and {Weinberg}, David H. and {Dav{\'e}}, Romeel},
        title = "{How do galaxies get their gas?}",
      journal = {\mnras},
         year = 2005,
        month = oct,
       volume = {363},
       number = {1},
        pages = {2-28},
          doi = {10.1111/j.1365-2966.2005.09451.x},
archivePrefix = {arXiv},
       eprint = {astro-ph/0407095},
 primaryClass = {astro-ph},
       adsurl = {https://ui.adsabs.harvard.edu/abs/2005MNRAS.363....2K}
}

@ARTICLE{Brooks09,
       author = {{Brooks}, A.~M. and {Governato}, F. and {Quinn}, T. and {Brook}, C.~B. and {Wadsley}, J.},
        title = "{The Role of Cold Flows in the Assembly of Galaxy Disks}",
      journal = {\apj},
         year = 2009,
        month = mar,
       volume = {694},
       number = {1},
        pages = {396-410},
          doi = {10.1088/0004-637X/694/1/396},
archivePrefix = {arXiv},
       eprint = {0812.0007},
 primaryClass = {astro-ph},
       adsurl = {https://ui.adsabs.harvard.edu/abs/2009ApJ...694..396B}
}

@ARTICLE{Chen20,
       author = {{Chen}, Hsiao-Wen and {Zahedy}, Fakhri S. and {Boettcher}, Erin and {Cooper}, Thomas M. and {Johnson}, Sean D. and {Rudie}, Gwen C. and {Chen}, Mandy C. and {Walth}, Gregory L. and {Cantalupo}, Sebastiano and {Cooksey}, Kathy L. and {Faucher-Gigu{\`e}re}, Claude-Andr{\'e} and {Greene}, Jenny E. and {Lopez}, Sebastian and {Mulchaey}, John S. and {Penton}, Steven V. and {Petitjean}, Patrick and {Putman}, Mary E. and {Rafelski}, Marc and {Rauch}, Michael and {Schaye}, Joop and {Simcoe}, Robert A. and {Weiner}, Benjamin J.},
        title = "{The Cosmic Ultraviolet Baryon Survey (CUBS) - I. Overview and the diverse environments of Lyman limit systems at z < 1}",
      journal = {\mnras},
         year = 2020,
        month = sep,
       volume = {497},
       number = {1},
        pages = {498-520},
          doi = {10.1093/mnras/staa1773},
archivePrefix = {arXiv},
       eprint = {2005.02408},
 primaryClass = {astro-ph.GA},
       adsurl = {https://ui.adsabs.harvard.edu/abs/2020MNRAS.497..498C}
}

@ARTICLE{Burchett19,
       author = {{Burchett}, Joseph N. and {Tripp}, Todd M. and {Prochaska}, J. Xavier and {Werk}, Jessica K. and {Tumlinson}, Jason and {Howk}, J. Christopher and {Willmer}, Christopher N.~A. and {Lehner}, Nicolas and {Meiring}, Joseph D. and {Bowen}, David V. and {Bordoloi}, Rongmon and {Peeples}, Molly S. and {Jenkins}, Edward B. and {O'Meara}, John M. and {Tejos}, Nicolas and {Katz}, Neal},
        title = "{The COS Absorption Survey of Baryon Harbors (CASBaH): Warm-Hot Circumgalactic Gas Reservoirs Traced by Ne VIII Absorption}",
      journal = {\apjl},
         year = 2019,
        month = jun,
       volume = {877},
       number = {2},
          eid = {L20},
        pages = {L20},
          doi = {10.3847/2041-8213/ab1f7f},
archivePrefix = {arXiv},
       eprint = {1810.06560},
 primaryClass = {astro-ph.GA},
       adsurl = {https://ui.adsabs.harvard.edu/abs/2019ApJ...877L..20B}
}

@ARTICLE{Werk16,
       author = {{Werk}, Jessica K. and {Prochaska}, J. Xavier and {Cantalupo}, Sebastiano and {Fox}, Andrew J. and {Oppenheimer}, Benjamin and {Tumlinson}, Jason and {Tripp}, Todd M. and {Lehner}, Nicolas and {McQuinn}, Matthew},
        title = "{The COS-Halos Survey: Origins of the Highly Ionized Circumgalactic Medium of Star-Forming Galaxies}",
      journal = {\apj},
         year = 2016,
        month = dec,
       volume = {833},
       number = {1},
          eid = {54},
        pages = {54},
          doi = {10.3847/1538-4357/833/1/54},
archivePrefix = {arXiv},
       eprint = {1609.00012},
 primaryClass = {astro-ph.GA},
       adsurl = {https://ui.adsabs.harvard.edu/abs/2016ApJ...833...54W}
}

@ARTICLE{Tumlinson17,
       author = {{Tumlinson}, Jason and {Peeples}, Molly S. and {Werk}, Jessica K.},
        title = "{The Circumgalactic Medium}",
      journal = {\araa},
         year = 2017,
        month = aug,
       volume = {55},
       number = {1},
        pages = {389-432},
          doi = {10.1146/annurev-astro-091916-055240},
archivePrefix = {arXiv},
       eprint = {1709.09180},
 primaryClass = {astro-ph.GA},
       adsurl = {https://ui.adsabs.harvard.edu/abs/2017ARA&A..55..389T}
}

@ARTICLE{Heckman90,
       author = {{Heckman}, Timothy M. and {Armus}, Lee and {Miley}, George K.},
        title = "{On the Nature and Implications of Starburst-driven Galactic Superwinds}",
      journal = {\apjs},
         year = 1990,
        month = dec,
       volume = {74},
        pages = {833},
          doi = {10.1086/191522},
       adsurl = {https://ui.adsabs.harvard.edu/abs/1990ApJS...74..833H}
}

@ARTICLE{Zheng2024,
       author = {{Zheng}, Yong and {Faerman}, Yakov and {Oppenheimer}, Benjamin D. and {Putman}, Mary E. and {McQuinn}, Kristen B.~W. and {Kirby}, Evan N. and {Burchett}, Joseph N. and {Telford}, O. Grace and {Werk}, Jessica K. and {Kim}, Doyeon A.},
        title = "{A Comprehensive Investigation of Metals in the Circumgalactic Medium of Nearby Dwarf Galaxies}",
      journal = {\apj},
         year = 2024,
        month = jan,
       volume = {960},
       number = {1},
          eid = {55},
        pages = {55},
          doi = {10.3847/1538-4357/acfe6b},
archivePrefix = {arXiv},
       eprint = {2301.12233},
 primaryClass = {astro-ph.GA},
       adsurl = {https://ui.adsabs.harvard.edu/abs/2024ApJ...960...55Z}
}

@ARTICLE{Nelson21,
       author = {{Nelson}, Dylan and {Byrohl}, Chris and {Peroux}, Celine and {Rubin}, Kate H.~R. and {Burchett}, Joseph N.},
        title = "{The cold circumgalactic medium in emission: Mg II haloes in TNG50}",
      journal = {\mnras},
         year = 2021,
        month = nov,
       volume = {507},
       number = {3},
        pages = {4445-4463},
          doi = {10.1093/mnras/stab2177},
archivePrefix = {arXiv},
       eprint = {2106.09023},
 primaryClass = {astro-ph.GA},
       adsurl = {https://ui.adsabs.harvard.edu/abs/2021MNRAS.507.4445N}
}

@ARTICLE{Saeedzadeh26,
       author = {{Saeedzadeh}, Vida and {Tumlinson}, Jason and {Peeples}, Molly S. and {O'Shea}, Brian W. and {Lochhaas}, Cassandra and {Corlies}, Lauren and {Trapp}, Cameron W. and {Smith}, Britton D. and {Werk}, Jessica K. and {Acharyya}, Ayan and {Augustin}, Ramona and {Lehner}, Nicolas and {Wright}, Anna C.},
        title = "{Figuring Out Gas \& Galaxies In Enzo (FOGGIE). XIV. The Observability of Emission from Accretion and Feedback in the Circumgalactic Medium with Current and Future Instruments}",
      journal = {\apj},
         year = 2026,
        month = jun,
       volume = {1004},
       number = {1},
          eid = {9},
        pages = {9},
          doi = {10.3847/1538-4357/ae658a},
archivePrefix = {arXiv},
       eprint = {2511.05644},
 primaryClass = {astro-ph.GA},
       adsurl = {https://ui.adsabs.harvard.edu/abs/2026ApJ..1004....9S}
}

@ARTICLE{Somerville_Dave15,
       author = {{Somerville}, Rachel S. and {Dav{\'e}}, Romeel},
        title = "{Physical Models of Galaxy Formation in a Cosmological Framework}",
      journal = {\araa},
         year = 2015,
        month = aug,
       volume = {53},
        pages = {51-113},
          doi = {10.1146/annurev-astro-082812-140951},
archivePrefix = {arXiv},
       eprint = {1412.2712},
 primaryClass = {astro-ph.GA},
       adsurl = {https://ui.adsabs.harvard.edu/abs/2015ARA&A..53...51S}
}

@ARTICLE{Piacitelli26,
       author = {{Piacitelli}, Daniel R. and {Brooks}, Alyson M. and {Sanchez}, N. Nicole and {Khatri}, Hetvi and {Christensen}, Charlotte and {Hummels}, Cameron and {Mishra}, Nishant and {Cruz}, Akaxia and {Keller}, Ben and {Quinn}, Thomas R. and {Shen}, Sijing and {Wadsley}, James},
        title = "{The Simulated Oxygen Shortage (SOS): Mapping the Missing OVI in Simulated Dwarf Galaxies to Subgrid Physics}",
      journal = {arXiv e-prints},
         year = 2026,
        month = may,
          eid = {arXiv:2605.15275},
        pages = {arXiv:2605.15275},
          doi = {10.48550/arXiv.2605.15275},
archivePrefix = {arXiv},
       eprint = {2605.15275},
 primaryClass = {astro-ph.GA},
       adsurl = {https://ui.adsabs.harvard.edu/abs/2026arXiv260515275P}
}

@ARTICLE{HWOCase,
       author = {{Burchett}, Joseph N. and {Lokhorst}, Deborah M. and {Faerman}, Yakov and {France}, Kevin and {Rubin}, Kate H.~R. and {Rupke}, David S.~N. and {Borthakur}, Sanchayeeta},
        title = "{Mapping galactic winds and small-scale structure in the circumgalactic medium with Habitable Worlds Observatory}",
      journal = {Journal of Astronomical Telescopes, Instruments, and Systems},
         year = 2025,
        month = oct,
       volume = {11},
          eid = {042213},
        pages = {042213},
          doi = {10.1117/1.JATIS.11.4.042213},
archivePrefix = {arXiv},
       eprint = {2507.03750},
 primaryClass = {astro-ph.GA},
       adsurl = {https://ui.adsabs.harvard.edu/abs/2025JATIS..11d2213B}
}

@ARTICLE{Ma16,
       author = {{Ma}, Xiangcheng and {Hopkins}, Philip F. and {Faucher-Gigu{\`e}re}, Claude-Andr{\'e} and {Zolman}, Nick and {Muratov}, Alexander L. and {Kere{\v{s}}}, Du{\v{s}}an and {Quataert}, Eliot},
        title = "{The origin and evolution of the galaxy mass-metallicity relation}",
      journal = {\mnras},
         year = 2016,
        month = feb,
       volume = {456},
       number = {2},
        pages = {2140-2156},
          doi = {10.1093/mnras/stv2659},
archivePrefix = {arXiv},
       eprint = {1504.02097},
 primaryClass = {astro-ph.GA},
       adsurl = {https://ui.adsabs.harvard.edu/abs/2016MNRAS.456.2140M}
}

@ARTICLE{Tremonti04,
       author = {{Tremonti}, Christy A. and {Heckman}, Timothy M. and {Kauffmann}, Guinevere and {Brinchmann}, Jarle and {Charlot}, St{\'e}phane and {White}, Simon D.~M. and {Seibert}, Mark and {Peng}, Eric W. and {Schlegel}, David J. and {Uomoto}, Alan and {Fukugita}, Masataka and {Brinkmann}, Jon},
        title = "{The Origin of the Mass-Metallicity Relation: Insights from 53,000 Star-forming Galaxies in the Sloan Digital Sky Survey}",
      journal = {\apj},
         year = 2004,
        month = oct,
       volume = {613},
       number = {2},
        pages = {898-913},
          doi = {10.1086/423264},
archivePrefix = {arXiv},
       eprint = {astro-ph/0405537},
 primaryClass = {astro-ph},
       adsurl = {https://ui.adsabs.harvard.edu/abs/2004ApJ...613..898T}
}

@ARTICLE{2026arXiv260718385Z,
       author = {{Zhang}, Huanian and {Xie}, Lizhi and {Qu}, Zhijie and {Zaritsky}, Dennis and {Ho}, Luis C. and {Bao}, Min and {Fontanot}, Fabio and {Hirschmann}, Michaela and {Liu}, Chenxu and {De Lucia}, Gabriella and {Wang}, Enci and {Zheng}, Haoyan},
        title = "{Star formation powers optical line emission from the CGM}",
      journal = {arXiv e-prints},
         year = 2026,
        month = jul,
          eid = {arXiv:2607.18385},
        pages = {arXiv:2607.18385},
          doi = {10.48550/arXiv.2607.18385},
archivePrefix = {arXiv},
       eprint = {2607.18385},
 primaryClass = {astro-ph.GA},
       adsurl = {https://ui.adsabs.harvard.edu/abs/2026arXiv260718385Z}
}

@ARTICLE{SciPy,
  author  = {Virtanen, Pauli and Gommers, Ralf and Oliphant, Travis E. and
            Haberland, Matt and Reddy, Tyler and Cournapeau, David and
            Burovski, Evgeni and Peterson, Pearu and Weckesser, Warren and
            Bright, Jonathan and {van der Walt}, St{\'e}fan J. and
            Brett, Matthew and Wilson, Joshua and Millman, K. Jarrod and
            Mayorov, Nikolay and Nelson, Andrew R. J. and Jones, Eric and
            Kern, Robert and Larson, Eric and Carey, C J and
            Polat, {\.I}lhan and Feng, Yu and Moore, Eric W. and
            {VanderPlas}, Jake and Laxalde, Denis and Perktold, Josef and
            Cimrman, Robert and Henriksen, Ian and Quintero, E. A. and
            Harris, Charles R. and Archibald, Anne M. and
            Ribeiro, Ant{\^o}nio H. and Pedregosa, Fabian and
            {van Mulbregt}, Paul and {SciPy 1.0 Contributors}},
  title   = {{{SciPy} 1.0: Fundamental Algorithms for Scientific
            Computing in Python}},
  journal = {Nature Methods},
  year    = {2020},
  volume  = {17},
  pages   = {261--272},
  adsurl  = {https://rdcu.be/b08Wh},
  doi     = {10.1038/s41592-019-0686-2},
}

@ARTICLE{Geda26,
       author = {{Geda}, Robel and {Cruz}, Akaxia and {Wright}, Anna C. and {Greene}, Jenny E. and {Brooks}, Alyson and {Quinn}, Thomas and {Wadsley}, James and {Keller}, Ben},
        title = "{The Formation of Dwarf Galaxy Disks}",
      journal = {\apj},
         year = 2026,
        month = jun,
       volume = {1004},
       number = {1},
          eid = {110},
        pages = {110},
          doi = {10.3847/1538-4357/ae64fd},
archivePrefix = {arXiv},
       eprint = {2510.26875},
 primaryClass = {astro-ph.GA},
       adsurl = {https://ui.adsabs.harvard.edu/abs/2026ApJ..1004..110G}
}

@ARTICLE{Munshi13,
       author = {{Munshi}, Ferah and {Governato}, F. and {Brooks}, A.~M. and {Christensen}, C. and {Shen}, S. and {Loebman}, S. and {Moster}, B. and {Quinn}, T. and {Wadsley}, J.},
        title = "{Reproducing the Stellar Mass/Halo Mass Relation in Simulated {\ensuremath{\Lambda}}CDM Galaxies: Theory versus Observational Estimates}",
      journal = {\apj},
         year = 2013,
        month = mar,
       volume = {766},
       number = {1},
          eid = {56},
        pages = {56},
          doi = {10.1088/0004-637X/766/1/56},
archivePrefix = {arXiv},
       eprint = {1209.1389},
 primaryClass = {astro-ph.CO},
       adsurl = {https://ui.adsabs.harvard.edu/abs/2013ApJ...766...56M}
}

@ARTICLE{Bellovary2010,
       author = {{Bellovary}, Jillian M. and {Governato}, Fabio and {Quinn}, Thomas R. and {Wadsley}, James and {Shen}, Sijing and {Volonteri}, Marta},
        title = "{Wandering Black Holes in Bright Disk Galaxy Halos}",
      journal = {\apjl},
         year = 2010,
        month = oct,
       volume = {721},
       number = {2},
        pages = {L148-L152},
          doi = {10.1088/2041-8205/721/2/L148},
archivePrefix = {arXiv},
       eprint = {1008.5147},
 primaryClass = {astro-ph.CO},
       adsurl = {https://ui.adsabs.harvard.edu/abs/2010ApJ...721L.148B}
}

@ARTICLE{scikit,
       author = {{van der Walt}, Stefan and {Sch{\"o}nberger}, Johannes L. and {Nunez-Iglesias}, Juan and {Boulogne}, Fran{\c{c}}ois and {Warner}, Joshua D. and {Yager}, Neil and {Gouillart}, Emmanuelle and {Yu}, Tony and {scikit-image Contributors}},
        title = "{scikit-image: Image processing in Python}",
      journal = {PeerJ},
         year = 2014,
        month = jan,
       volume = {2},
          eid = {e453},
        pages = {e453},
          doi = {10.7717/peerj.453},
archivePrefix = {arXiv},
       eprint = {1407.6245},
 primaryClass = {cs.MS},
       adsurl = {https://ui.adsabs.harvard.edu/abs/2014PeerJ...2..453V}
}

@ARTICLE{Corlies20,
       author = {{Corlies}, Lauren and {Peeples}, Molly S. and {Tumlinson}, Jason and {O'Shea}, Brian W. and {Lehner}, Nicolas and {Howk}, J. Christopher and {O'Meara}, John M. and {Smith}, Britton D.},
        title = "{Figuring Out Gas \& Galaxies in Enzo (FOGGIE). II. Emission from the z = 3 Circumgalactic Medium}",
      journal = {\apj},
         year = 2020,
        month = jun,
       volume = {896},
       number = {2},
          eid = {125},
        pages = {125},
          doi = {10.3847/1538-4357/ab9310},
archivePrefix = {arXiv},
       eprint = {1811.05060},
 primaryClass = {astro-ph.GA},
       adsurl = {https://ui.adsabs.harvard.edu/abs/2020ApJ...896..125C}
}

@ARTICLE{Bordoloi11,
       author = {{Bordoloi}, R. and {Lilly}, S.~J. and {Knobel}, C. and {Bolzonella}, M. and {Kampczyk}, P. and {Carollo}, C.~M. and {Iovino}, A. and {Zucca}, E. and {Contini}, T. and {Kneib}, J.-P. and {Le Fevre}, O. and {Mainieri}, V. and {Renzini}, A. and {Scodeggio}, M. and {Zamorani}, G. and {Balestra}, I. and {Bardelli}, S. and {Bongiorno}, A. and {Caputi}, K. and {Cucciati}, O. and {de la Torre}, S. and {de Ravel}, L. and {Garilli}, B. and {Kova{\v{c}}}, K. and {Lamareille}, F. and {Le Borgne}, J.-F. and {Le Brun}, V. and {Maier}, C. and {Mignoli}, M. and {Pello}, R. and {Peng}, Y. and {Perez Montero}, E. and {Presotto}, V. and {Scarlata}, C. and {Silverman}, J. and {Tanaka}, M. and {Tasca}, L. and {Tresse}, L. and {Vergani}, D. and {Barnes}, L. and {Cappi}, A. and {Cimatti}, A. and {Coppa}, G. and {Diener}, C. and {Franzetti}, P. and {Koekemoer}, A. and {L{\'o}pez-Sanjuan}, C. and {McCracken}, H.~J. and {Moresco}, M. and {Nair}, P. and {Oesch}, P. and {Pozzetti}, L. and {Welikala}, N.},
        title = "{The Radial and Azimuthal Profiles of Mg II Absorption around 0.5 < z < 0.9 zCOSMOS Galaxies of Different Colors, Masses, and Environments}",
      journal = {\apj},
         year = 2011,
        month = dec,
       volume = {743},
       number = {1},
          eid = {10},
        pages = {10},
          doi = {10.1088/0004-637X/743/1/10},
archivePrefix = {arXiv},
       eprint = {1106.0616},
 primaryClass = {astro-ph.CO},
       adsurl = {https://ui.adsabs.harvard.edu/abs/2011ApJ...743...10B}
}

@ARTICLE{Peroux20,
       author = {{P{\'e}roux}, C{\'e}line and {Nelson}, Dylan and {van de Voort}, Freeke and {Pillepich}, Annalisa and {Marinacci}, Federico and {Vogelsberger}, Mark and {Hernquist}, Lars},
        title = "{Predictions for the angular dependence of gas mass flow rate and metallicity in the circumgalactic medium}",
      journal = {\mnras},
         year = 2020,
        month = dec,
       volume = {499},
       number = {2},
        pages = {2462-2473},
          doi = {10.1093/mnras/staa2888},
archivePrefix = {arXiv},
       eprint = {2009.07809},
 primaryClass = {astro-ph.GA},
       adsurl = {https://ui.adsabs.harvard.edu/abs/2020MNRAS.499.2462P}
}

@ARTICLE{Tchernyshyov23,
       author = {{Tchernyshyov}, Kirill and {Werk}, Jessica K. and {Wilde}, Matthew C. and {Prochaska}, J. Xavier and {Tripp}, Todd M. and {Burchett}, Joseph N. and {Bordoloi}, Rongmon and {Howk}, J. Christopher and {Lehner}, Nicolas and {O'Meara}, John M. and {Tejos}, Nicolas and {Tumlinson}, Jason},
        title = "{The CGM$^{2}$ Survey: Quenching and the Transformation of the Circumgalactic Medium}",
      journal = {\apj},
         year = 2023,
        month = jun,
       volume = {949},
       number = {2},
          eid = {41},
        pages = {41},
          doi = {10.3847/1538-4357/acc86a},
archivePrefix = {arXiv},
       eprint = {2211.06436},
 primaryClass = {astro-ph.GA},
       adsurl = {https://ui.adsabs.harvard.edu/abs/2023ApJ...949...41T}
}

@ARTICLE{Lochhaas26,
       author = {{Lochhaas}, Cassandra and {Peeples}, Molly S. and {O'Shea}, Brian W. and {Tumlinson}, Jason and {Corlies}, Lauren and {Saeedzadeh}, Vida and {Lehner}, Nicolas and {Wright}, Anna C. and {Werk}, Jessica K. and {Trapp}, Cameron W. and {Augustin}, Ramona and {Acharyya}, Ayan and {Smith}, Britton D. and {Vargas}, Carlos J.},
        title = "{Figuring Out Gas \& Galaxies in Enzo (FOGGIE). XI. Circumgalactic O VI Emission Traces Clumpy Inflowing Recycled Gas}",
      journal = {\apj},
         year = 2026,
        month = mar,
       volume = {1000},
       number = {1},
          eid = {104},
        pages = {104},
          doi = {10.3847/1538-4357/ae458f},
archivePrefix = {arXiv},
       eprint = {2510.25844},
 primaryClass = {astro-ph.GA},
       adsurl = {https://ui.adsabs.harvard.edu/abs/2026ApJ..1000..104L}
}

@ARTICLE{Tumlinson11,
       author = {{Tumlinson}, J. and {Thom}, C. and {Werk}, J.~K. and {Prochaska}, J.~X. and {Tripp}, T.~M. and {Weinberg}, D.~H. and {Peeples}, M.~S. and {O'Meara}, J.~M. and {Oppenheimer}, B.~D. and {Meiring}, J.~D. and {Katz}, N.~S. and {Dav{\'e}}, R. and {Ford}, A.~B. and {Sembach}, K.~R.},
        title = "{The Large, Oxygen-Rich Halos of Star-Forming Galaxies Are a Major Reservoir of Galactic Metals}",
      journal = {Science},
         year = 2011,
        month = nov,
       volume = {334},
       number = {6058},
        pages = {948},
          doi = {10.1126/science.1209840},
archivePrefix = {arXiv},
       eprint = {1111.3980},
 primaryClass = {astro-ph.CO},
       adsurl = {https://ui.adsabs.harvard.edu/abs/2011Sci...334..948T}
}

@ARTICLE{Chen09,
       author = {{Chen}, Hsiao-Wen and {Mulchaey}, John S.},
        title = "{Probing The Intergalactic Medium-Galaxy Connection At z < 0.5. I. A Galaxy Survey In Qso Fields And A Galaxy-Absorber Cross-Correlation Study}",
      journal = {\apj},
         year = 2009,
        month = aug,
       volume = {701},
       number = {2},
        pages = {1219-1242},
          doi = {10.1088/0004-637X/701/2/1219},
archivePrefix = {arXiv},
       eprint = {0906.3293},
 primaryClass = {astro-ph.CO},
       adsurl = {https://ui.adsabs.harvard.edu/abs/2009ApJ...701.1219C}
}

@ARTICLE{Bradford15,
       author = {{Bradford}, Jeremy D. and {Geha}, Marla C. and {Blanton}, Michael R.},
        title = "{A Study in Blue: The Baryon Content of Isolated Low-mass Galaxies}",
      journal = {\apj},
         year = 2015,
        month = aug,
       volume = {809},
       number = {2},
          eid = {146},
        pages = {146},
          doi = {10.1088/0004-637X/809/2/146},
archivePrefix = {arXiv},
       eprint = {1505.04819},
 primaryClass = {astro-ph.GA},
       adsurl = {https://ui.adsabs.harvard.edu/abs/2015ApJ...809..146B}
}

@ARTICLE{Huang12,
       author = {{Huang}, Shan and {Haynes}, Martha P. and {Giovanelli}, Riccardo and {Brinchmann}, Jarle},
        title = "{The Arecibo Legacy Fast ALFA Survey: The Galaxy Population Detected by ALFALFA}",
      journal = {\apj},
         year = 2012,
        month = sep,
       volume = {756},
       number = {2},
          eid = {113},
        pages = {113},
          doi = {10.1088/0004-637X/756/2/113},
archivePrefix = {arXiv},
       eprint = {1207.0523},
 primaryClass = {astro-ph.CO},
       adsurl = {https://ui.adsabs.harvard.edu/abs/2012ApJ...756..113H}
}

@ARTICLE{Lin25,
       author = {{Lin}, Zeren and {Martin}, D. Christopher and {Matuszewski}, Mateusz and {Neill}, James D. and {Miles}, Drew M. and {Picouet}, Vincent and {Prusinski}, Nikolaus Z. and {Hoadley}, Keri and {McGurk}, Rosalie},
        title = "{Kinematically Complex Circumgalactic Gas Around a Low-mass Galaxy: Filamentary Inflow and Counterrotation in J0910b}",
      journal = {\apj},
         year = 2025,
        month = dec,
       volume = {995},
       number = {1},
          eid = {12},
        pages = {12},
          doi = {10.3847/1538-4357/ae10b2},
       adsurl = {https://ui.adsabs.harvard.edu/abs/2025ApJ...995...12L}
}

@ARTICLE{Dutta24_musestack,
       author = {{Dutta}, Rajeshwari and {Fumagalli}, Michele and {Fossati}, Matteo and {Rafelski}, Marc and {Revalski}, Mitchell and {Arrigoni Battaia}, Fabrizio and {D'Odorico}, Valentina and {P{\'e}roux}, Celine and {Prichard}, Laura J. and {Swinbank}, A. Mark},
        title = "{Metal line emission around z < 1 galaxies}",
      journal = {\aap},
         year = 2024,
        month = nov,
       volume = {691},
          eid = {A236},
        pages = {A236},
          doi = {10.1051/0004-6361/202450733},
archivePrefix = {arXiv},
       eprint = {2409.02182},
 primaryClass = {astro-ph.GA},
       adsurl = {https://ui.adsabs.harvard.edu/abs/2024A&A...691A.236D}
}

@ARTICLE{Nielsen24,
       author = {{Nielsen}, Nikole M. and {Fisher}, Deanne B. and {Kacprzak}, Glenn G. and {Chisholm}, John and {Martin}, D. Christopher and {Reichardt Chu}, Bronwyn and {Sandstrom}, Karin M. and {Rickards Vaught}, Ryan J.},
        title = "{An emission map of the disk-circumgalactic medium transition in starburst IRAS 08339+6517}",
      journal = {Nature Astronomy},
         year = 2024,
        month = dec,
       volume = {8},
        pages = {1602-1609},
          doi = {10.1038/s41550-024-02365-x},
archivePrefix = {arXiv},
       eprint = {2311.00856},
 primaryClass = {astro-ph.GA},
       adsurl = {https://ui.adsabs.harvard.edu/abs/2024NatAs...8.1602N}
}

@article{Abraham_2014,
   title={Ultra–Low Surface Brightness Imaging with the Dragonfly Telephoto Array},
   volume={126},
   ISSN={1538-3873},
   url={http://dx.doi.org/10.1086/674875},
   DOI={10.1086/674875},
   number={935},
   journal={Publications of the Astronomical Society of the Pacific},
   publisher={IOP Publishing},
   author={Abraham, Roberto G. and van Dokkum, Pieter G.},
   year={2014},
   month=Jan, pages={55–69} }

@ARTICLE{Chen25,
       author = {{Chen}, Seery and {Lokhorst}, Deborah M. and {Pasha}, Imad and {Bowman}, William P. and {Janssens}, Steven R. and {Rhea}, Carter and {Liu}, Qing and {Shen}, Zili and {Abraham}, Roberto G. and {van Dokkum}, Pieter},
        title = "{First Light with the 120-lens Dragonfly Spectral Line Mapper}",
      journal = {\pasp},
         year = 2025,
        month = aug,
       volume = {137},
       number = {8},
          eid = {084103},
        pages = {084103},
          doi = {10.1088/1538-3873/adedb8},
       adsurl = {https://ui.adsabs.harvard.edu/abs/2025PASP..137h4103C} }

@ARTICLE{Lokhorst24,
       author = {{Lokhorst}, Deborah M. and {Chen}, Seery and {Pasha}, Imad and {Purcell}, Victoria and {Bowman}, William P. and {Liu}, Qing and {Shen}, Zili and {MacNichol}, Aidan and {Malakhov}, Evgeni I. and {Abraham}, Roberto G. and {van Dokkum}, Pieter},
        title = "{Realizing the potential of the Dragonfly Spectral Line Mapper: Calibration methods and on-sky performance}",
      journal = {arXiv e-prints},
         year = 2024,
        month = jun,
          eid = {arXiv:2406.17979},
        pages = {arXiv:2406.17979},
          doi = {10.48550/arXiv.2406.17979},
archivePrefix = {arXiv},
       eprint = {2406.17979},
 primaryClass = {astro-ph.IM},
       adsurl = {https://ui.adsabs.harvard.edu/abs/2024arXiv240617979L}
}

@ARTICLE{Pasha24,
       author = {{Pasha}, Imad and {Chen}, Seery and {Lokhorst}, Deborah and {Bowman}, William P. and {Shen}, Zili and {Liu}, Qing and {Malakhov}, Evgeni I. and {Abraham}, Roberto and {van Dokkum}, Pieter G.},
        title = "{Software infrastructure for the highly-distributed semi-autonomous Dragonfly Spectral Line Mapper}",
      journal = {arXiv e-prints},
         year = 2024,
        month = jun,
          eid = {arXiv:2406.15301},
        pages = {arXiv:2406.15301},
          doi = {10.48550/arXiv.2406.15301},
archivePrefix = {arXiv},
       eprint = {2406.15301},
 primaryClass = {astro-ph.IM},
       adsurl = {https://ui.adsabs.harvard.edu/abs/2024arXiv240615301P}
}

@inproceedings{Chung21b,
author = {Haeun Chung and Carlos J. Vargas and Erika Hamden and Tom McMahon and Kerry Gonzales and Aafaque R. Khan and Simran Agarwal and Hop Bailey and Peter Behroozi and Trenton Brendel and Heejoo Choi and Tom Connors and Lauren Corlies and Jason Corliss and Ralf-J{\"u}rgen Dettmar and David Dolana and Ewan S. Douglas and John Guzman and Dave Hamara and Walt Harris and Karl Harshman and Carl Hergenrother and Keri Hoadley and John Kidd and Daewook Kim and Jessica S. Li and Manny Montoya and Corwynn Sauve and David Schiminovich and Sanford Selznick and Oswald Siegmund and Michael Ward and Ellie M. Wolcott and Dennis Zaritsky},
title = {{Aspera: the UV SmallSat telescope to detect and map the warm-hot gas phase in nearby galaxy halos}},
volume = {11819},
booktitle = {UV/Optical/IR Space Telescopes and Instruments: Innovative Technologies and Concepts X},
editor = {Allison A. Barto and James B. Breckinridge and H. Philip Stahl},
organization = {International Society for Optics and Photonics},
publisher = {SPIE},
pages = {1181903},
year = {2021},
doi = {10.1117/12.2593001},
URL = {https://doi.org/10.1117/12.2593001}
}

@book{burger2010principles,
  title={Principles of Digital Image Processing: Fundamental Techniques},
  author={Burger, W. and Burge, M.J.},
  isbn={9781848001916},
  lccn={2008942779},
  series={Undergraduate Topics in Computer Science},
  url={https://books.google.com/books?id=2LIMMD9FVXkC},
  year={2010},
  publisher={Springer London}
}

@ARTICLE{Pillepich18,
       author = {{Pillepich}, Annalisa and {Springel}, Volker and {Nelson}, Dylan and {Genel}, Shy and {Naiman}, Jill and {Pakmor}, R{\"u}diger and {Hernquist}, Lars and {Torrey}, Paul and {Vogelsberger}, Mark and {Weinberger}, Rainer and {Marinacci}, Federico},
        title = "{Simulating galaxy formation with the IllustrisTNG model}",
      journal = {\mnras},
         year = 2018,
        month = jan,
       volume = {473},
       number = {3},
        pages = {4077-4106},
          doi = {10.1093/mnras/stx2656},
archivePrefix = {arXiv},
       eprint = {1703.02970},
 primaryClass = {astro-ph.GA},
       adsurl = {https://ui.adsabs.harvard.edu/abs/2018MNRAS.473.4077P}
}

@ARTICLE{Booth09,
       author = {{Booth}, C.~M. and {Schaye}, Joop},
        title = "{Cosmological simulations of the growth of supermassive black holes and feedback from active galactic nuclei: method and tests}",
      journal = {\mnras},
         year = 2009,
        month = sep,
       volume = {398},
       number = {1},
        pages = {53-74},
          doi = {10.1111/j.1365-2966.2009.15043.x},
archivePrefix = {arXiv},
       eprint = {0904.2572},
 primaryClass = {astro-ph.CO},
       adsurl = {https://ui.adsabs.harvard.edu/abs/2009MNRAS.398...53B}
}
\bibliographystyle{aasjournalv7}

\end{document}